\documentclass{aa}   
\usepackage{epsfig,psfig,natbib,txfonts}
\bibpunct{(}{)}{;}{a}{}{,} 
\newcommand{\kms}{km\,s$^{-1}$\,}
\newcommand{\Apixel}{\AA\,pixel$^{-1}$\, }
\newcommand{\arcpixel}{arcsec pixel$^{-1}$\,}

\newcommand{\HI}{$(H\!+\!I)$}
\begin{document}

%
   \title{Gas flow and dark matter in the inner parts of early-type
   barred galaxies}

   \subtitle{I. SPH simulations and comparison with the observed
   kinematics}

   \author{I. P\'erez  \inst{1,2} \and R. Fux\inst{2,3} \and
           K. Freeman\inst{2} }

   \offprints{I. P\'erez Mart\'{\i}n \\ email: isa@astro.rug.nl}

   \institute{Kapteyn Astronomical Institute, University of Groningen,
   Postbus 800, Groningen 9700 AV, Netherlands \and RSAA, Mt. Stromlo
   and Siding Spr. Obs., Private Bag, Woden PO, Canberra, ACT 2606,
   Australia \and  Geneva Observatory, Ch. des Maillettes 51, CH-1290
   Sauverny, Switzerland }

   \date{}

   \abstract{

This paper presents the dynamical simulations run in the potential
derived from the light distribution of 5 late-type barred spiral
galaxies (\object{IC~5186}, \object{NGC~5728}, \object{NGC~7267},
\object{NGC~7483} and \object{ NGC~5505}). The aim is to determine
whether the mass distribution together with the hydrodynamical
simulations can reproduce the observed line-of-sight velocity curves and the gas
morphology in the inner regions of the sample barred galaxies. The
light distribution is obtained from the $H$-band and the $I$-band
combined together. The M/L is determined using population synthesis
models. The observations and the methodology of the mass distribution
modelling are presented in a companion paper.  The SPH models using
the stellar mass models obtained directly from the $H$-band light
distributions give a good representation of the gas distribution and
dynamics of the modelled galaxies, supporting the maximum disk
assumption. This result indicates that the gravitational field in the
inner region is mostly provided by the stellar luminous
component. When 40\% of the total mass is transferred to an
axisymmetric dark halo, the modelled kinematics clearly depart from
the observed kinematics, whereas the departures are negligible for
dark mass halos of 5\% and 20\% of the total mass. This result sets a
lower limit for the contribution of the luminous component of about
80\%, which is in agreement with the maximum disk definition of the
stellar mass contribution to the rotation curve (about
85\%$\pm$10). This result is in agreement with the results found by
\citet{weiner01} for \object{NGC~4123} using a similar
methodology. For two galaxies, NGC 7483 and IC 5186, a very good
agreement with the observed data is found. In these cases the non
circular motions can help to break the disk-halo degeneracy. For the
other three galaxies (NGC~5728, NGC~7267 and NGC~5505) no definite
results are found: for NGC~7267 and NGC~5505 no steady state is
reached in the simulations and for NGC~5728 there is no good agreement
with the observed kinematics, possibly due to the presence of a
secondary bar decoupled from the primary. However, for this latter
galaxy the M/L ratio used gives the right amplitude of the rotation
curve, in further support of the M/L calculation method used
throughout this work. Fast bars give the best fit to the observed
kinematics for NGC~7483 and IC 5186 with corotation at the end of the
bar for NGC~7483 and at 1.4$\times$~$R_{\rm bar}$  for IC~5186. For
NGC~5505 for  which no steady state configuration is found, the
addition of a rigid halo stabilises the gas flows but the derived
kinematics does not fit well the observations.

       } \maketitle

%

\section{Introduction}

Flat rotation curves offer the most direct evidence for the existence
of dark matter in galaxies. The simplest interpretation is that spiral
galaxies possess massive dark halos that extend to larger radii than
the optical disks.  However, in order to get this flat rotation curve
both components need to be coupled in some way.  This coupling leads
to a degenerate problem;  the velocity field of an axisymmetric galaxy
does not allow to uniquely disentangle the contribution of the halo
from that of the disk.  This disk-halo degeneracy requires a further
constraint, usually the assumption of ``the maximum disk''. A galaxy
is described as having a ``maximum disk'' when the stellar disk
accounts for most of the rotational support of the galaxy in the inner
parts.  A quantitative definition for the maximum disk hypothesis is
that the stellar disk provides 85\% $\pm$ 10\% of the total rotational
support of the galaxy at $R_{\rm disk}$=2.2$h_{\rm R}$, where $h_{\rm
R}$ is the disk scale-length \citep{sackett97}. The upper boundary
avoids hollow cores in the dark halo.  There is no consensus yet on
whether maximum disks are preferred
\citep{freeman92,courteau99,salucci99}. The match to the overall shape
of the rotation curve is often met just as well by pure halo or pure
disk models. It is not even clear whether the Milky Way favours a
maximum disk or not.\\

Gas dynamics within bars can help to break this degeneracy. The
non-circular velocities that characterise gas streaming motions in a
bar will allow us to determine the necessity for a dark halo in the
inner parts of spiral galaxies.  It has been argued that a heavy halo
will slow down a bar by dynamical friction
\citep{weinberg85,debattista98}; therefore the bars embedded in a heavy
halo would have to be slow rotators. However, the few directly
measured pattern speeds of bars seem to indicate that they are fast
rotators, implying the non existence of a dominant halo in the inner
parts. \cite*{tremaine} explain the fact that the bars rotate fast by
arguing that the angular momentum transfer from the disk to the halo
would spin the latter up, flattening the inner halo. Since bars are
coupled to the halos they should also rotate rapidly. In this way one
obtains a fast bar embedded in a dark halo; see also
\citet{athanassoula02} for bar-halo coupling results.\\

 To test whether the luminous mass in the inner parts of spiral
galaxies can account for their observed gas kinematics or whether a
more axisymmetric dark matter component is required, the inner gas
dynamics of a sample of barred spiral galaxies has been model led by
running a 3-D composite N-body/hydrocode. The complicated shapes of
barred galaxies have led us to a non-analytical approach to the
modelling of the gas dynamics in these systems and the potential is
calculated directly from the light distribution.  This approach takes
into account the non-axisymmetric structures, such as bars, that are
very important in the dynamics of the inner region.  Some similar
previous work can be found in \cite*{weiner01}.\\

Section~\ref{data} briefly presents the observation, data reduction
and calibration carried out.  In Section~\ref{introductiontothecode}
the code used for modelling the gas flow will be presented. How the
mass distribution is obtained from the composite $H$ and $I$ band
light distribution is explained in Section~\ref{massdistribution}. The
initial conditions for the simulations will be presented in
Section~\ref{initialconditions}. The parameter space explored in the
simulations is presented in Section~\ref{modelparameters}. In Section
~\ref{thebestfitmodels} the analysis methods used to compare the
modelled and observed line-of-sight velocity curves will be
introduced. In Section ~\ref{ic5186} and Section~\ref{resultsic5186}
the results for IC~5186 are presented. This section also presents
simulations within the potential derived from the $I$-band alone and
compares them to those based on the composite \HI-band image
(Section~\ref{modelwithmassdistributionderivedfromtheibandlightdistribution}).
In Section~\ref{comparisonofthemodelsfordifferentscaleheights} the
comparison for IC~5186 for different scale-heights is presented. In
Section~\ref{ngc5728} the results for NGC 5728 are
shown. Section~\ref{ngc7267} presents the results for
NGC~7267. Section~\ref{ngc7483} shows the results for NGC
7483. Section~\ref{ngc5505} presents the results for NGC
5505. Section~\ref{modelswithadarkhalocomponent} presents the results
of the modelling with a dark halo component, which is done for
IC~5186 and NGC~5505. A discussion is given in
Section~\ref{discussionchapter5} and Section~\ref{conclusion}
concludes the paper.

\section{Data}
\label{data}
The modelled galaxies form part of a larger sample of 27 isolated
spiral galaxies. The whole data set, data analysis and sample
characterization is presented in  P\'erez et al. (in prep., hereafter
'data' paper) and chapter 2 of P\'erez PhD. thesis \citeyearpar{perez3}.

\subsection {Optical rotation curves}
\label{opticalrotationcurvesobservations}

The rotation curves were obtained by optical spectroscopic (H$\alpha$)
observations along several position angles using the Cassegrain Boller
\& Chivens spectrograph on the 1.52m telescope at La Silla Observatory
and the Double Beam Spectrograph (DBS) on the 2.3m telescope at Siding
Spring Observatory (SSO). At La Silla the detector was a 2K$\times$2K
Loral-Lesser CCD, and the dispersion was 1.0 \Apixel (grating
\#26). The pixel scale of the CCD is 0.82 arcsec pixel$^{-1}$, the
wavelength range studied was $\lambda$ 4900--5700 \AA. The detector at
SSO was a SITe $1752\times532$ CCD. The gratings used were the 1200B
and 1200R for the blue and red arms respectively, giving a dispersion
of 1.13 \Apixel for the blue arm and 1.09 \Apixel for the red arm in
the wavelength interval from 4315--5283 \AA\ and 6020--6976 \AA\
respectively. The slit was in both cases visually aligned with the
galaxy nucleus at different position angles (PA) and the slit width
was set to 1.5 arcsec on the sky, giving a projected width of 2.2
pixels.\\

Total integration times for the galaxies varied from 1 to 3 hours. In
the case of the DBS the exposures had to be shorter than 20 minutes
due to flexure of the instrument. The observations were split in three
or more exposures per galaxy to eliminate cosmic rays from the
frames.\\

All the spectra obtained were reduced using standard IRAF
routines. Overscan and bias were subtracted. No dark subtraction was
done due to the low dark current of the chips used (1$e^{-}$
pix$^{-1}$ hr$^{-1}$). At the 1.52m telescope, dome and sky flats were
acquired in order to correct properly for slit width
variation. Flatfielding correction was achieved to the 2\%
level. Geometrical corrections were applied to the frames in order to
correct for any  misalignment. The data reduction for the DBS spectra
was carried out in a similar way but, due to the double beam nature of
the spectrograph, care had to be taken to deconvolve the dichroic
using spectrophotometric standards.\\

The ionised gas rotation curve and the velocity dispersions were
obtained by fitting a Gaussian to H$\alpha$ and [N{\small II}]
($\lambda$6583.4\AA) lines using the MIDAS package ALICE. These lines
were clearly detected in all the galaxies observed. The gas rotation
curves and the velocity dispersion profile derived independently from
H$\alpha$ and [N{\small II}] are in good agreement at all radii.

\subsection{Optical imaging}
\label{opticalimaging}

Optical imaging observations were performed at the Swope 40 inch
telescope at Las Campanas Observatory (LCO) and the 40 inch telescope
at SSO. The observations at the Swope telescope were taken with SITe 3
chip with a pixel scale of 0.435 \arcpixel and a field of view of 12
arcmin. The observations at SSO were taken with the direct imager and
the SITe 2K detector with 0.6 \arcpixel. $B$, $V$ and $I$ filters were
used, and three images were taken per object per band, except in some
cases where only two frames were obtained. Three standard fields
~\citep{landolt} were acquired per night. A series of twilight sky
flats were obtained each night.\\

The optical data were reduced using standard IRAF routines. Bias and
overscan subtraction were performed. The bias levels were very stable
during the runs. For the SSO observations the CCDs had to be
reinitialised during the night to maintain stable and structure-free
bias frames.\\

Calibration for the photometric nights was done using APPHOT on the
Landolt fields and then using PHOTCAL to fit the zero-point
magnitudes, the colour and the extinction coefficients.

\subsection {Near-Infrared imaging}
\label{infraredimagingobservations}

Near-Infrared (NIR) photometry gives a good representation of the old
stellar population and therefore of the stellar mass
distribution. Furthermore, IR bands are less affected by the internal
absorption of the galaxy.\\

NIR imaging data were obtained at LCO,  using the DuPont 100 inch
telescope and the Swope 40 inch telescope, and the 2.3m telescope at
SSO. The NIR observations at the DuPont were obtained with the IR
camera with a Rockwell NICMOS3 HgCdTe  256$\times$256 array with 0.42
\arcpixel\ (1.8 arcmin $\times$ 1.8 arcmin field of view). The IR
observations at the Swope telescope were taken with the IR camera with
a NICMOS3 HgCdTe 256$\times$256 with a pixel scale of 0.599 \arcpixel\
(2.5 arcmin $\times$ 2.5 arcmin field of view). CASPIR, the imager at
the 2.3m telescope at SSO, uses a Santa Barbara Research Center
256$\times$256 InSb detector array, with a focal plane scale of 0.5
\arcpixel. The filters used were $J$,$H$,$K_{\rm n}$ for the imaging
with CASPIR and $J_{\rm short}$,$H$,$K_{\rm short}$ for the LCO
runs. The NICMOS3 detector becomes nonlinear when the total counts
(sky + object) exceed 17,000 ADU. For this reason and  the sky
variability in the IR, we exposed for 60--120 s in $J$ 30 s in $H$ and
15 s in $K$. Each object was observed at several positions on the
array. \\

For the NIR imaging obtained at LCO only the smallest angular size
galaxies of the sample were observed. Therefore no offset to different
sky positions was made and the sky was simply calculated by averaging
a number of the stacked images for a given object with a $\sigma$
clipping rejection algorithm.\\

The objects measured with CASPIR were recorded at several positions on
the array with a dither pattern and larger offsets were applied to
obtain sky frames with the same exposure time as the object
frames. The dithering between frames was typically 20 arcsec and the
offset for the sky frames was typically of 2.5 arcmin. Sky frames were
obtained after each second object exposure, beginning and ending each
exposure sequence with a sky exposure. Each single exposure time was
the same as for the LCO runs.\\

The NIR data was reduced using standard IRAF routines and {\it
eclipse} (IR data reduction package developed at ESO).  All the frames
had to be corrected for instrumental effects such as nonlinearity,
dark current and pixel-to-pixel variation. The response of the CASPIR
detector has a quadratic nonlinearity which is corrected after bias
subtraction but before dark correction and flatfielding.\\

The HST standards list ~\citep{persson} and the IRIS standard star list
~\citep{carter} were used. 6 to 7 standard star observations were
observed per night, whenever possible repeating each standard up to 3
times. Photometric calibration was carried out in a similar way to the
optical imaging calibration.

\section{Introduction to the code}
\label{introductiontothecode}

The N-body and hydro code was initially developed by the Geneva
Observatory galactic dynamics group for spiral galaxies studies
\citep{pfenniger93,fux97,fux99}.  This code has been modified by Fux
for the purpose of the present work.  The main difference with the
original code used by Fux in his thesis \citep{fux97} is that his
simulations were completely self-consistent; here the stellar
potential is fixed using the observed light distribution. Only a
single grid was used for the potential and force calculations, since
there was no addition of a vertically extended halo.  What follows is
a brief summary of the code characteristics. For more details about
the code, refer to \cite{fux99} or his PhD thesis, \cite{fux97}.\\

The code uses the particle-mesh technique to assign to each cell of a
fixed grid a mass that is proportional to the enclosed number of
particles in that cell, according to the Cloud-In-Cell method (CIC).
Then the potential at each of the grid points is calculated from the
masses at all grid points. In our simulations, the rigid potential is
computed only once, assigning a Monte Carlo realisation of the mass
distribution derived from the galaxy images to the grid. The CIC
method is a method to linearly interpolate a discretely sampled field
where each particle is represented by a 'cloud' with the same volume
as that of the grid cell and the mass is split up between the eight
cells the cloud can cover. Fractions of that 'particle-cloud' mass are
then assigned to the grid points and the sum over all the particles
gives the mass per cell. The potential is then calculated by Fourier
transforming the density and the adopted kernel, multiplying both and
then inverse Fourier transform in the $z$ and $\phi$ dimension to
recover the potential (in the $R$ dimension, the potential does not
appear as a convolution and therefore a direct summation is
performed). The geometry chosen is a cylindrical-polar grid and the
short range forced are softened using a variable homogeneous ellipsoid
kernel with principal axes matched to the local grid resolution in
each dimension.  The polar grid gives a better radial resolution at
the central parts of the disk where the density of matter is higher
and the rotation period shorter.\\

The pressure and viscous forces are derived by 3-D Smooth Particle
Hydrodynamics, SPH ( e.g. \citep{benz90,fux99}).  In all
the models the gas is taken to be isothermal with a sound speed of 10
kms$^{-1}$ and an adiabatic index of 5/3, corresponding to the atomic
hydrogen.\\

The gas particle positions and velocities are advanced by integrating
the equations of motion using an integrator similar to the leap-frog
algorithm but evaluating the phase-space coordinates at the same
times. The code uses an adaptative time step. After each integration
step, one evaluates the maximum relative contribution per unit time of
the second order terms to the integrated magnitudes. If this maximum
contribution exceeds a given tolerance, then the code restores the old
values and integrates again with a smaller time step. For a constant
time step, the integrator is time reversible and the precision remains
close to that of the leap-frog. The adaptative time step is very
useful to resolve the shocks in the gas.

\section{Simulations}
\label{simulations}

The strategy in running the simulations was the following: first, low
resolution simulations were run for all the galaxies. This was to
optimise the different parameters, such as the onset time for the
non-axisymmetric component, the vertical grid size, the total
integration time and also whether running the code with no gas
self-gravity, i.e considering the gas as test particles, makes a
significant difference to the final results. These simulations were
run at the Research School of Astronomy and Astrophysics (RSAA) to
ensure no waste of time when running the high resolution simulations
at the supercomputer (see below).\\

The size of the grid for these low resolution simulations is 31 cells
in the radial direction, 32 in the tangential direction and 242
(including doubling up cells) in the vertical direction with a total
of 50,000 particles. The rotation direction of the bar is chosen to be
opposite to the winding of the spiral arms outside the bar. Only two
pattern speeds were checked (placing corotation (CR) at the end of the
bar and at twice the bar semi-major axis) and six onset times for the
non-axisymmetric component were tried, with the bar fully grown at
0.5, 1.0, 2.0, 3.0, 4.0 and 5.0 bar rotations. These simulations were
carried out to ensure that at the chosen onset time the particle flow
adjusted steadily (see also Sect.~\ref{initialconditions}). \\

The high resolution simulations were run at the Australian National
University (ANU) supercomputer facility using the Alpha Server SC
system. Several models were run using $H$ and $I$ photometry for the
different galaxies to explore the bar pattern speed parameter space in
order to reproduce the observed  position-velocity diagrams.  Then,
for IC~5186 several runs with the best fit bar pattern speed were
carried out but this time using for the mass distribution only the
$I$-band instead of the composite \HI-band image. This was done to
test the effect of using a band that is more affected by
extinction. The results of this test are presented in Section
~\ref{modelwithmassdistributionderivedfromtheibandlightdistribution}.\\

In order to check whether the observed kinematics could be reproduced
with an additional axisymmetric component, we ran simulations with a
dark halo component on two of the five galaxies
(Section~\ref{modelswithadarkhalocomponent}).
In practice,
 this is done in the following way.  When growing the bar in the
 beginning of a simulation, the gravitational potential is decomposed
 into \begin{equation} \Phi~=f~\Phi_{\rm full}~+~(1~\!-~\!f)~\Phi_{\rm
 axisym}, \end{equation} where $\Phi_{\rm full}$ is the full
 non-axisymmetric potential as derived from the population synthesis
 models, $\Phi_{\rm axisym}$ its azimuthal average, and $f$ a factor
 that increases linearly with time from 0 to 1 during the onset
 time. If one stops the growth of the bar at a percentage $f$ of that
 onset time and then carries on the simulation maintaining this partly
 grown bar, the effect is such as replacing a fraction $(1\!-\!f)$ of
 the inferred visible mass by an axisymmetric component, which we
 consider here as our dark matter halo. In this way, we reduce
 the azimuthal forces while keeping the same average circular
 velocity. Furthermore, this method, which reduces the strength of the
 bar while leaving the total mass and the axisymmetric part of the
 rotation curve unchanged, was very easy to implement in the
 code. The flatness of this dark halo it is not a crucial problem
 in the modelling because the gas dynamics do not depend much on the
 vertical forces. \\

The gas flow in the inner parts will be influenced by the choice of
the vertical scale-height, with a higher scale-height giving
effectively a smoother potential. Tests with different scale-heights
were carried out for IC~5186 to investigate how the computed
kinematics are affected (see
Section~\ref{comparisonofthemodelsfordifferentscaleheights}). \\

The high resolution simulations were run with the following
 characteristics: grid size used is 95 cells in the radial direction,
 96 in the azimuthal and 1214 (including doubling up) in the vertical
 directions. The vertical resolution is set to 0.05 times the
 scale-height adopted for the luminous mass distribution. The number
 of gas particles used is 300,000. The barred potential rotates at
 fixed pattern speed $\Omega_{b}$. Low resolution simulations show gas
 self-gravity to be irrelevant for the results so we finally ran the
 code with no self-gravity.

\subsection{Mass distribution}
\label{massdistribution}

The gravitational field is calculated from a surface density
distribution that follows the light distribution of the combined $H$
and $I$-band images of the projected to face-on image, as described in
P\'erez et al. (in prep., hereafter paper II) and chapters 3 and 4 of
P\'erez PhD. thesis \citeyearpar{perez3}. The M/L ratio is fixed using
stellar population synthesis models as described also in paper II. In
this way, we also test the reliability of the population synthesis M/L
ratios. In a manner consistent with previous works \citep{jong96}, where colour gradients in the disks of spiral galaxies are
found, 2-D M/L ratio maps taking into account these colour changes
were built, assuming that they are mainly due to population
changes. In the process of calculating the M/L ratios it is clear that
the main factor affecting the values obtained is the Initial Mass
Function (IMF). We adopted a Salpeter IMF with a flatter slope for \(
$m$ < 0.5\) $M_{\odot}$. Though one might think that this is a
degenerate problem where we cannot distinguish between a scenario with
a different IMF and that of a sub-maximum disk, in fact the M/L
adopted is not only a factor since we are assuming a 2-D M/L map and
we are not only looking at the maximum of the rotation curve, but also
the shape of the line-of-sight (L-O-S) velocity curves.\\

For the deprojection of the galaxies a flat disk was adopted and a
simple geometrical deprojection was applied. An analytical model was
chosen for the extinction. In this model the stars lie above and below
the layer of the gas and dust. A radially exponential varying optical
depth was added to the model. The methodology followed is described in
paper II.\\

For the vertical distribution an exponential profile is assumed with a
radially constant vertical scale-height. This may not be a good
assumption as there is evidence from infrared photometric studies of
the Milky Way that the vertical scale of the bar is larger than that
of the disk \citep{freudenreich98}. The biggest impact on the dynamics
of a non-constant scale-height is precisely in the inner region we are
interested in, where the radial forces will change
significantly. However, it is hard to do better than this since not
much is known about the scale-height of bars in external galaxies. For
most of the simulations one value of the scale-height was adopted,
following the relationship found by \cite{kregel02}.  They analysed
the structure of the stellar disk in a sample of edge-on  galaxies and
found that the average $<$$h_{\rm R}$/$h_{\rm z}$$>$=7.3$\pm$2.2,
where h$_{\rm R}$ and $h_{\rm z}$ are respectively the disk
exponential scale-length and scale-height. They also found an average
$R_{\rm max}$/$h_{\rm R}$=3.6$\pm$ 0.6, where R$_{\rm max}$ represents
the disk truncation radius. The scale-lengths are derived from the
$H$-band images. The different values of the scale-height and the
scale-length in the simulations are presented in
Table~\ref{tab:parameters}. Table~\ref{tab:galaxies} presents the gas
and stellar masses for reference although the gas mass is not included
in the potential of the simulations because we do not know its
distribution. The gas mass represents the HI mass; no information on
CO is found in the literature for most of the galaxies, apart from
NGC~5728 \citep{combes02} where the CO mass represents
3.0$\times$10$^{9}$~M$_{\odot}$, comparable to the HI mass. Even when
adding the CO mass to the HI mass the total gas mass still represents
a small fraction of the total stellar mass.
\begin{table}
\begin{center}
\caption{HI gas and stellar masses for the modelled galaxies}
\label{tab:galaxies}
\vspace{5mm}
\begin{tabular}{lll}
\hline\hline Name& HI mass in M$_{\odot}$ & Stellar mass in
M$_{\odot}$\\ \hline IC
5186&9.36$\times$10$^{9}$&4.94$\times$10$^{10}$\\ NGC
7483&6.49$\times$10$^{9}$ &2.37$\times$10$^{11}$\\ NGC
5505&2.01$\times$10$^{9}$ & 1.16$\times$10$^{11}$\\ NGC
5728&2.5$\times$10$^{9}$ & 4.70$\times$10$^{11}$\\ NGC
7267&1.00$\times$10$^{9}$ & 6.86$\times$10$^{10}$\\ \hline
\end{tabular}
\end{center}
\end{table}

\subsection{Initial conditions}
\label{initialconditions}

The initial gas density in the simulation consists of one
component. The radial distribution for the gas is a Beta function with
a standard deviation set to the scale-length of the visible disk
(Table~\ref{tab:parameters}) and radially vanishing at a distance 4
times this scale-length. The vertical distribution of the gas is
generated directly by solving the hydrostatic equilibrium equation for
an isothermal gas. Different initial density distributions have little
effect on the results. The gas particles have pure circular motion
with cylindrical rotation and  zero velocity dispersion, since the
effective dispersion is taken into account in the pressure component
of the SPH.\\

Based on the low resolution simulations, the bar is chosen to grow
linearly during three bar rotations, so that the gas flow can steadily
adjust to the forcing bar without requiring to much CPU time. For
growing times shorter than this the flow took some more time to
settle, depending on the galaxy, and of course always reached the same
final state. However, if one chooses a shorter bar growth time and
does not run the simulations long enough one could be seeing
misleading transient features and interpret them as part of the steady
flow.

\subsection{Model parameters}
\label{modelparameters}

We made a series of standard runs for each galaxy, and then some
exploratory runs to test the effect of varying different parameters in
the simulations.  The standard runs were run with fixed stellar M/L
ratio and fixed constant scale-height.  Several bar parameters have a
large effect on the gas response: the central concentration, the axial
ratio, the mass of the bar and the pattern speed. The first three are
fixed by the given potential (in the standard runs)  and the only one
that can be varied is the pattern speed, $\Omega_{\rm b}$. This
parameter fixes the position of the resonances and controls much of
the gas response.\\

For each galaxy 6 different pattern speeds were modelled. The
$\Omega_{\rm b}$ is parametrised by the ratio of the corotation radius
assuming axisymmetry to the bar semi-major axis $R_{\rm CR}/R_{\rm
bar}$. Table ~\ref{tab:gal_par} give the
pattern speeds modeled for the different galaxies. The bar radius has
been obtained as explained in the data paper.\\

For the test runs the pattern speed was fixed to that giving the best
model (see Section~\ref{thebestfitmodels} for an explanation of the
criteria) in the standard runs and other parameters that could affect
the gas response were explored.  The scale-height of the stellar
component has also an effect on the dynamics of the gas in the inner
parts.  Larger scale-heights smooth out the potential, reducing the
radial gas flows.  For IC~5186, the scale-height dependence of the
velocity flow was investigated. Models with three different scale
heights at three pattern speeds were run.  The scale-height explored
were 0.5, 1.0 and 1.5 times the one used in the standard runs.

For the models with a rigid halo (IC~5186 and NGC~5505): 5, 20 and
40\% of the total mass ascribed to the dark halo in the manner
explained in Section ~\ref{simulations} were tested.

\begin{table}
\centering
\caption{Exponential scale-height and scale-lengths, distance and grid
size for the modelled galaxies}
\begin{center}
\label{tab:parameters}
\vspace{5mm}
\begin{tabular}{lcccc}
\hline\hline  Name & Distance& $h_{\rm R}$&  $h_{\rm z}$& Grid
extent\\ &(Mpc)&(kpc)&(kpc)&(kpc)\\ \hline IC 5186& 65.57 &   4.591 &
0.629 & 55.0 \\ NGC 7483 &  65.85  & 11.00 &  1.507 &  104.0\\ NGC
5505 & 57.13  & 1.897 &  0.260  &  29.0 \\ NGC 5728 &  37.17  & 5.451
& 0.747 &  90.0 \\ NGC 7267 & 44.70  &  2.503 &  0.343 &  41.0\\
\hline\hline
\end{tabular}
\end{center}
\end{table}

\subsection{The best fit models}
\label{thebestfitmodels}

The best fit model requires in the first place that the model slit
position-velocity diagrams best reproduce the observed rotation curve
with the following criteria. The position-velocity diagrams are
extracted from the models for comparison with the long-slit
observations.  The velocities are projected on the plane of the sky,
inclining the model as the real galaxy.  Then, a virtual slit is
placed on the model and the position-velocity diagram is extracted at
the observed position angle. The kinematics of the SPH particles
are not convolved with the SPH kernel this fact, however, does not
affect the derivation of the first velocity moments. To compare the
modelled and observed position-velocity diagrams, the simulated data
is linearly interpolated in position to the observed data.  Then a
$\chi^{2}$ comparison of the two velocity profiles is computed. 
When computing the $\chi^{2}$ a velocity dispersion component of
10\kms\ is added, corresponding to the assumed sound speed in the
models.  Because our aim is to exploit the constraining power of the
non-circular motions induced by the bar, we restrict the comparison to
the bar region only, where these motions are strongest, and exclude
the surrounding disk. Furthermore, we are assuming the same pattern
speed for the spiral arms as for the bar; since these are non
self-consistent models and the spiral arms are more likely to be
transient features, it is safest to consider only the bar region for
our comparison with the observed kinematics.\\

It is hard to compare the galaxies' gas distribution with the modelled
gas density distribution since no 2-D information on the gas
distribution for the sample galaxies exist in the literature. In order
to get an idea of how good the modelled gas density distribution is, a
comparison with the observed images is performed. The $B$-band light
follows mainly the regions dominated by young stellar population,
indicating regions of recent star formation. These regions are in turn
tracers of gas and dust in the galaxy. The dust lanes can be used to
trace shocks in the gas flow \citep{athanassoula92} which can be
compared to the modelled velocity fields. The dust lanes are
correlated with high densities in the gas distribution. Visual
comparison of the modelled gas distribution with the $B$-band images
was performed. In order to emphasise the non-axysimmetric part of the
gas distribution of the non-axysimmetric part a masking technique is
applied to both the observed galaxies and the modelled
distribution. An axisymmetric distribution of the $B$-band image was
created by averaging azimuthally the flux over ellipses centered on
the galaxy at fixed position angle and ellipticity and then subtracted
from the original image. The same was done to the modelled density map
convolved with a Gaussian to imitate the $B$-band image spatial
resolution. Now we discuss the individual galaxies.

\section{IC 5186}
\label{ic5186}

\begin{figure}
\vspace{0cm} \hbox{\hspace{0cm}\epsfxsize=4.5cm
\epsfbox{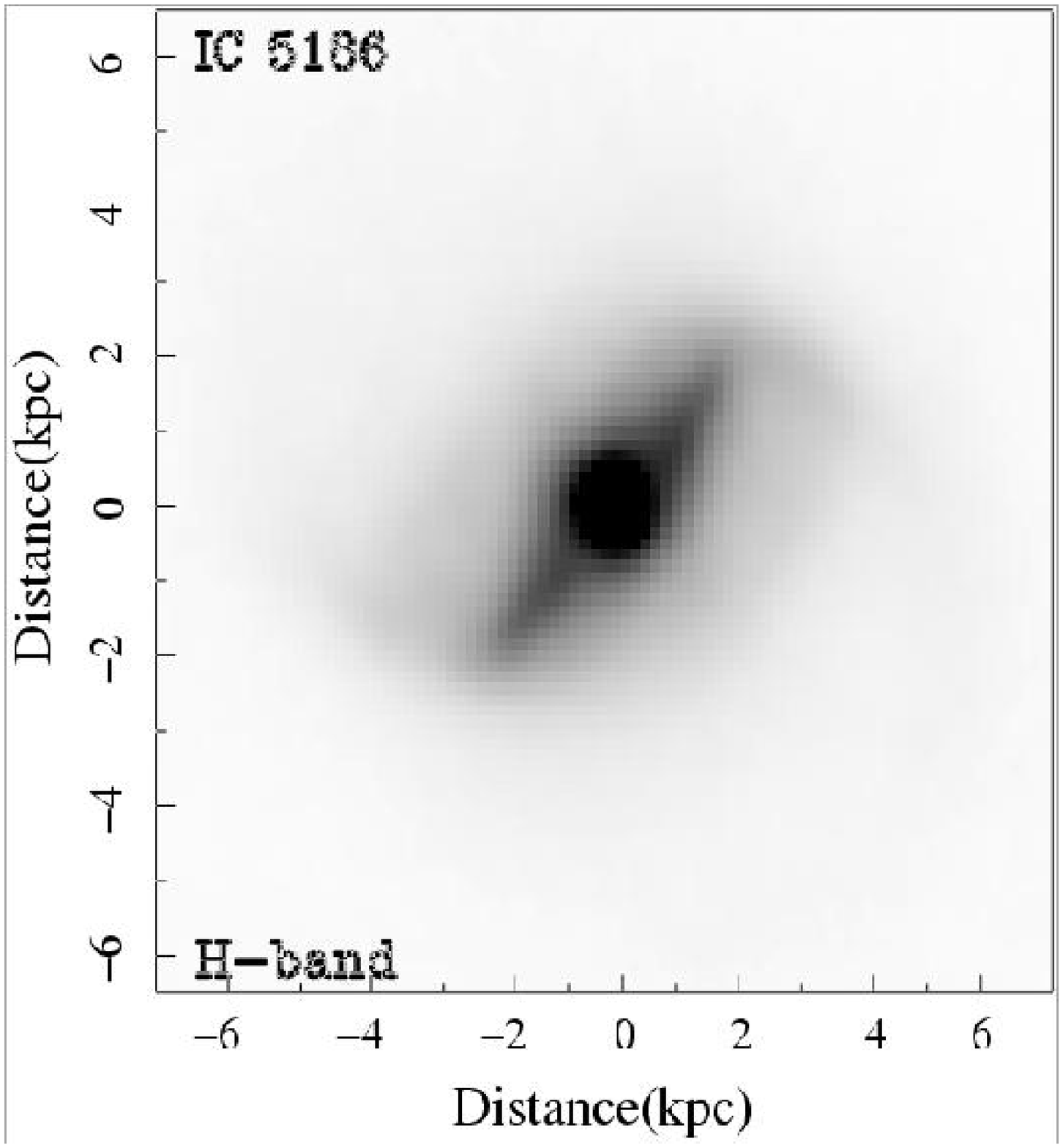} \epsfxsize=4.45cm
\epsfbox{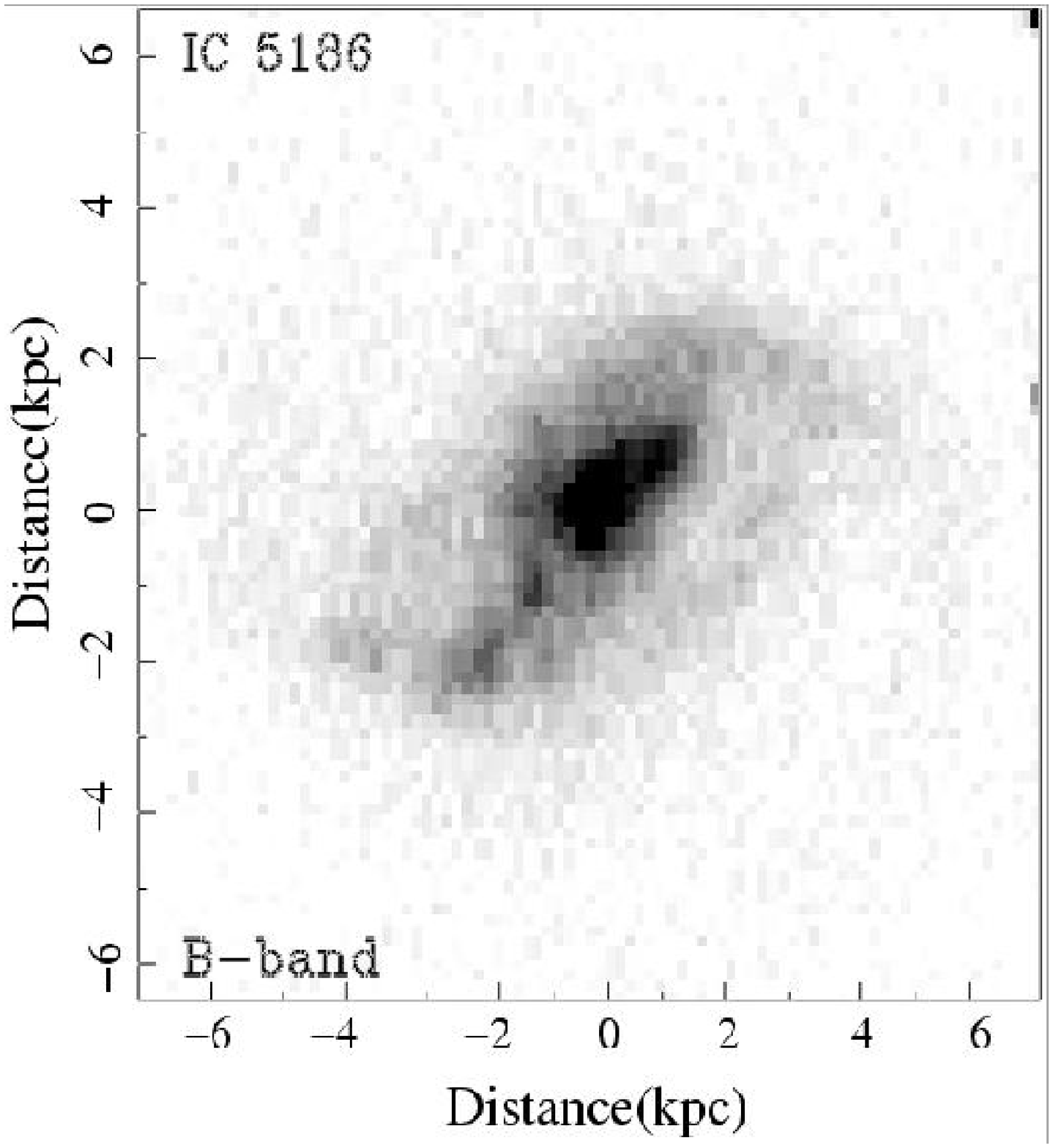}}
\vspace{0cm}
\caption[Light distribution in different bands for IC~5186]{Light
distribution in different bands for IC~5186. The left panel shows the
light distribution in the $H$-band and the right panel the light
distribution in the $B$-band. Notice the smoother distribution of the
light in the inner region for the $H$-band.}
\label{fig:light}
\end{figure}

IC 5186 is an isolated SBab galaxy, at a distance of 65 Mpc.  There is
no kinematic nor photometric study of this galaxy in the literature.
Optically an inner ring, a bar and filamentary structure in the region
between the bar and the ring are observed.  The $K$ and $H$-band
images show a much smoother distribution of the inner region (see
Fig.~\ref{fig:light}).  Long slit spectra along the major axis (when
speaking about the major/minor axis kinematics this always refers
throughout the text to the apparent principal axes of the galaxy
projected on the sky) shows a steep gradient in the inner part of the
rotation curve with a maximum velocity of 180~\kms\ at 2 arcsec.  The
minor axis shows a double component. Although it is difficult to
disentangle the different components, there seems to be a velocity
difference between the components of $\approx$ 200 km s$^{-1}$ over a
region of about 2.5 kpc (data paper). The bar position angle is
$\approx$ 45$^{\circ}$ with respect to the line of nodes.\\

The optical emission line spectra show [N{\small
II}]/H$\alpha$~$\approx$~1 which could suggest shocks; however, no
[O{\small I}] ($\lambda$ 6300$\AA$)  is present and $\log$~([S{\small
II}]/H$\alpha$)~$\approx$~-1.0 (typical of HII regions).  The
ionisation temperature is probably low since no trace of [O{\small
III}]$\lambda$ 5007$\AA$ is found in any of the spectra.  The
observations show that $\log$(O/H) is high, probably higher than
solar, since O{\small III}/H$\beta$ is around 1/3, which is consistent
with abundances found in the central parts of other galaxies.

\subsection{Results}
\label{resultsic5186}

\begin{table}
\begin{center}
\caption{Parameters for the test runs on IC~5186}
\label{tab:ic5186_test}
\vspace{5mm}
\begin{tabular}{lcc}
\hline\hline $R_{\rm CR}$/$R_{\rm bar}$ & scale-height of & \% of dark
halo mass \\ &of stellar disk (kpc) &from the total mass\\ \hline 1.0&
0.629& 5\\ 1.0&0.629&20\\ 1.0&0.629&40\\ 1.0&0.943&0\\ 1.0&0.314&0\\
\hline 1.4& 0.629& 5\\ 1.4&0.629&20\\ 1.4&0.629&40\\ 1.4&0.943&0\\
1.4&0.314&0\\ \hline 1.6& 0.629& 5\\ 1.6&0.629&20\\ 1.6&0.629&40\\
1.6&0.943&0\\ 1.6&0.314&0\\ \hline\hline
\end{tabular}
\end{center}
\end{table}

After three bar rotations, when the non-axisymmetric component is
fully grown, the gas distribution of the simulations in the potential
derived from the composite \HI-band light distribution settles in a
steady configuration for all the bar pattern speeds. In all cases an
elongated inner ring forms parallel to the bar major axis. Inside this
ring for $R_{\rm CR}$/$R_{\rm bar}$~$<$~1.4 a nuclear spiral is
formed but for $R_{\rm CR}$/$R_{\rm bar}$~$\geq$~1.4 a nuclear ring
forms with no mini spiral structure. From the frequency plot for the
axisymmetric system no inner Lindblad resonaces (ILRs) are
present. This nuclear ring is elongated parallel to the major axis of
the bar and does not seem to be related to any resonance. To get a
proper insight on the relationship of the different features with the
different periodic orbits, an orbit analysis is needed.\\

\begin{center}
\begin{figure*}
\vspace{0cm}
\hspace{0cm}\psfig{figure=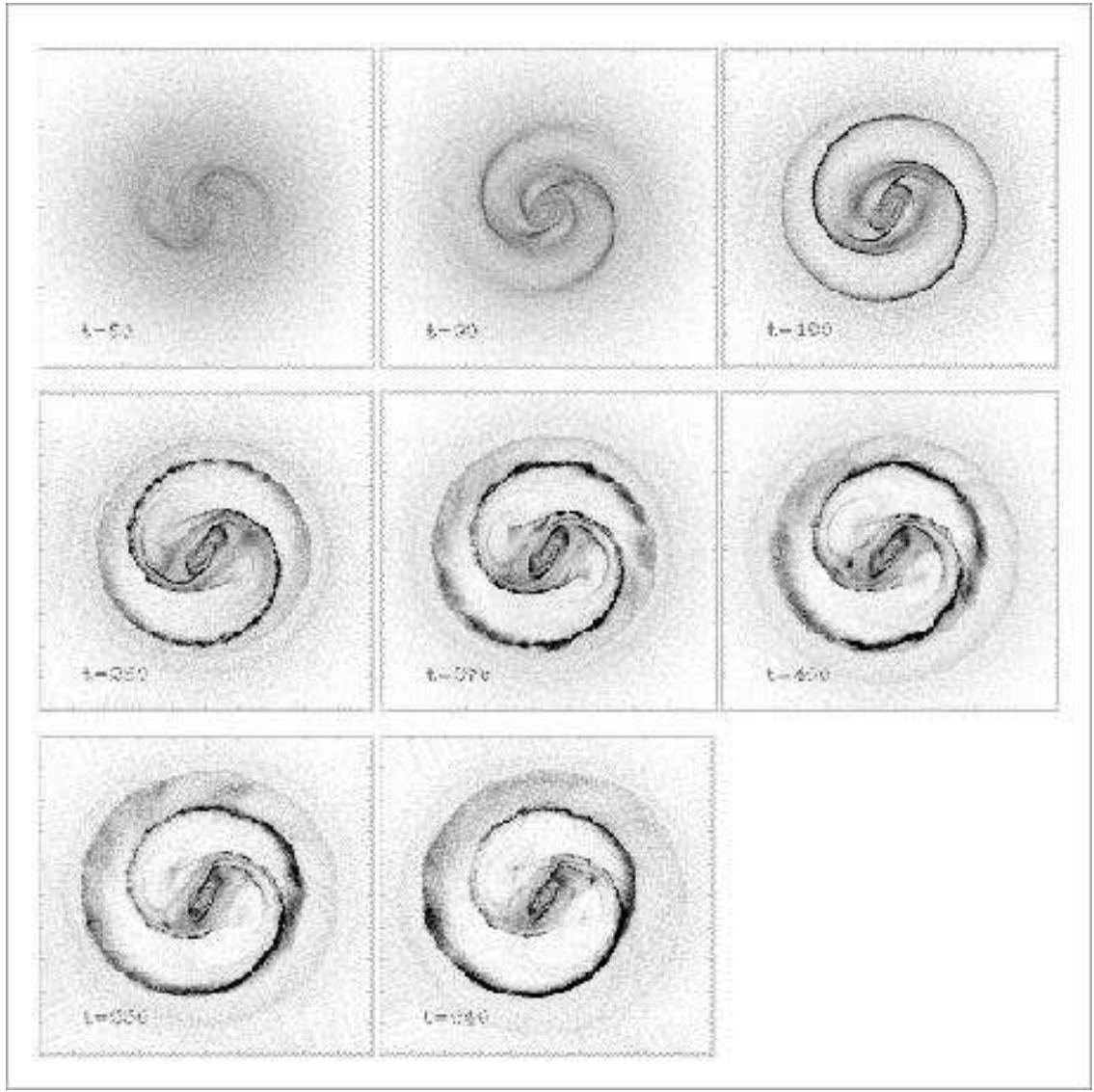,width=15.0cm,angle=-90}
\vspace{0cm}
\caption[Time evolution of the gas distribution of IC5186]{Time
evolution of the gas distribution in the IC5186 simulation with
corotation at 1.4$\times$the bar semi-major axis. Each panel, from
left to right and top to bottom, shows the distribution at different
integration times, from 1 bar rotation (second panel) to 7 bar
rotations, and in the frame corotating with the bar. The
non-axisymmetric component is fully grown after three bar rotations
and the bar pattern is rotating counter-clockwise in the inertial
frame. The size of the frames is 20 kpc in diameter and the time is in
Myr. Notice the gas trapped around the L$_{4,5}$ points after three
bar rotations.}
\label{fig:ic5186evolgas}
\end{figure*}
\end{center}

\begin{figure*}
\begin{center}
\vspace{0cm}
\hspace{0cm}\psfig{figure=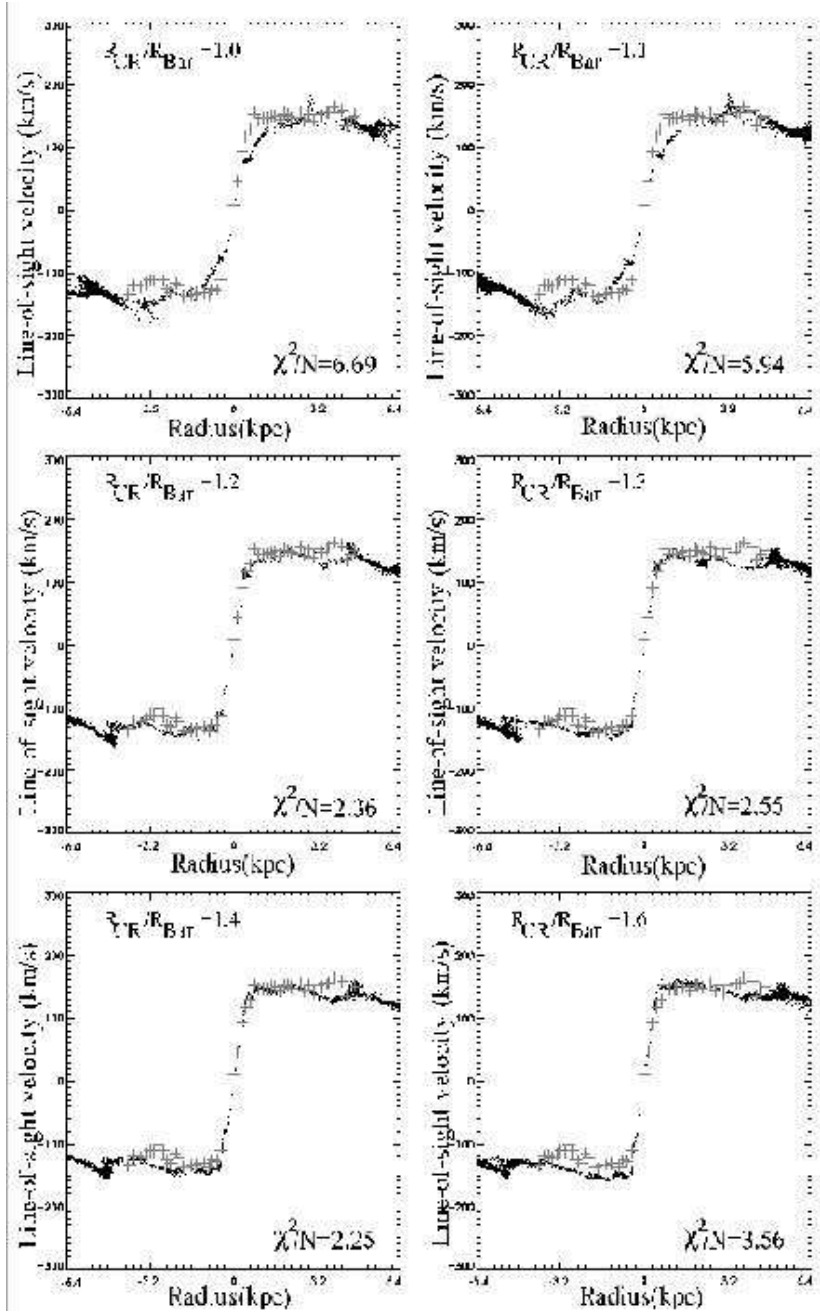,width=11.0cm}
\vspace{0cm}
\caption[Position-velocity diagrams along the major axis for IC~5186]{
Line-of-sight velocity curves along the major axis for the IC~5186 models
with different bar pattern speeds, after 8 bar rotations. The dots
represent the gas particles and the overlaid gray crosses the observed
major axis line-of-sight velocity curve. The $R_{\rm CR}/R_{\rm bar}$ for each
curve is indicated in the panels. The best fit to
the observed data corresponds to a $R_{\rm CR}/R_{\rm bar}$=1.4}
\label{fig:vel1}
\end{center}
\end{figure*}

Morphologically there is little change in the gas distribution for the
different pattern speeds, apart from these very central features and
the fact that the features move outward as the pattern speed
decreases. Fig.~\ref{fig:ic5186evolgas} shows the time evolution of
IC~5186 for a pattern speed of $\Omega_{\rm b}$=63.4 km s$^{-1}$
kpc$^{-1}$ ($R_{\rm CR}$/$R_{\rm bar}$=1.4); each figure shows the gas
density evolution after a further bar rotation. At all times, gas
seems to be trapped in orbits around the $L_{4}$ and the $L_{5}$
Lagrangian points, near to the minor axis of the
bar. Fig.~\ref{fig:vel1} shows the line-of-sight velocity curves along
the major axis for models with $R_{\rm CR}$/$R_{\rm bar}$ of 1.0, 1.1,
1.2, 1.3, 1.4 and 1.6. From this figure it is not immediately obvious
which one is the best fit. However, $R_{\rm CR}$/$R_{\rm bar}$ =1.4
looks best supported by the $\chi^{2}$ statistic, presented in
Table~\ref{tab:gal_par}. The $\chi^{2}$ computed for IC~5186
includes only the major axis data.\\

\begin{figure*}
\begin{center}
\vspace{0cm}
\hspace{0cm}\psfig{figure=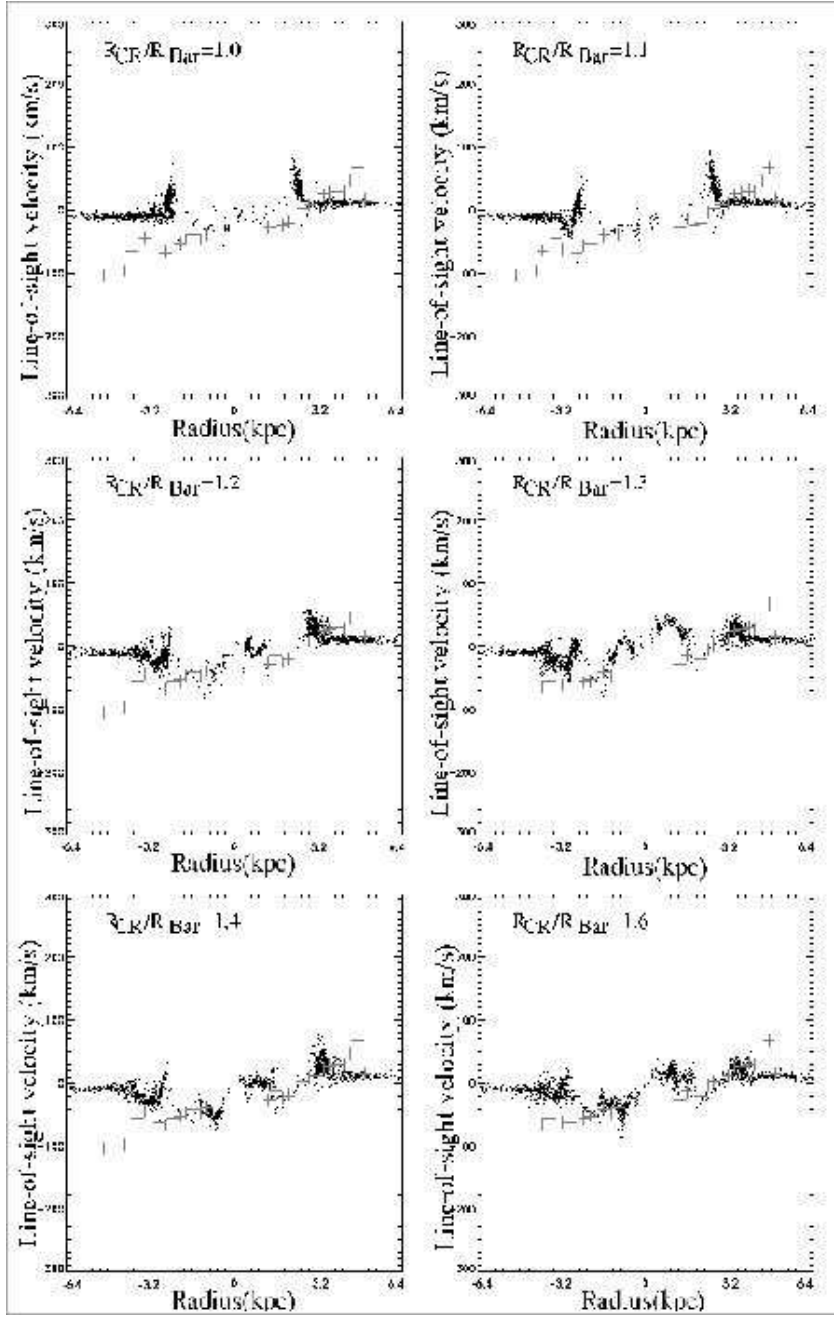,width=11.0cm}
\vspace{0cm}
\caption[Position-velocity diagrams along the minor axis for
IC~5186]{Line-of-sight velocity curves along the minor axis for models
with different bar pattern speeds for the IC~5186 simulations after 8
bar rotations . The overlaid gray crosses represent the observed minor axis
line-of-sight velocity curve and the dots represent the gas
particles. The $R_{\rm CR}/R_{\rm bar}$ for each curve is indicated in
the panels. Outside the central 10 arcsec the motions in the models
became circular, whereas non-circular motions are still present in the
observed data. Although difficult to interpret, the best fit seems to
be the model with $R_{\rm CR}/R_{\rm bar}$=1.4}
\label{fig:ic5186min}
\end{center}
\end{figure*}

For the minor axis, as already mentioned, the emission lines show very
broad profiles with a hardly distinguishable double
component. Therefore, the position-velocity diagram is more difficult
to interpret.  In Fig.~\ref{fig:ic5186min} the modelled minor axis
kinematics is plotted together with the data. The central emission
lines of the observed data  have not been plotted due to the fact that
no single line profile could be fitted. The model with corotation at
1.4 times the bar semi-major axis is consistent with the observed
kinematics.  The model shows highly non-circular motions on the minor
axis, only within  the bar region, whereas in the observations the
non-circular motions extend beyond the bar region.\\
    
\begin{figure*}
\begin{center}
\vspace{0cm}
\hspace{0cm}\psfig{figure=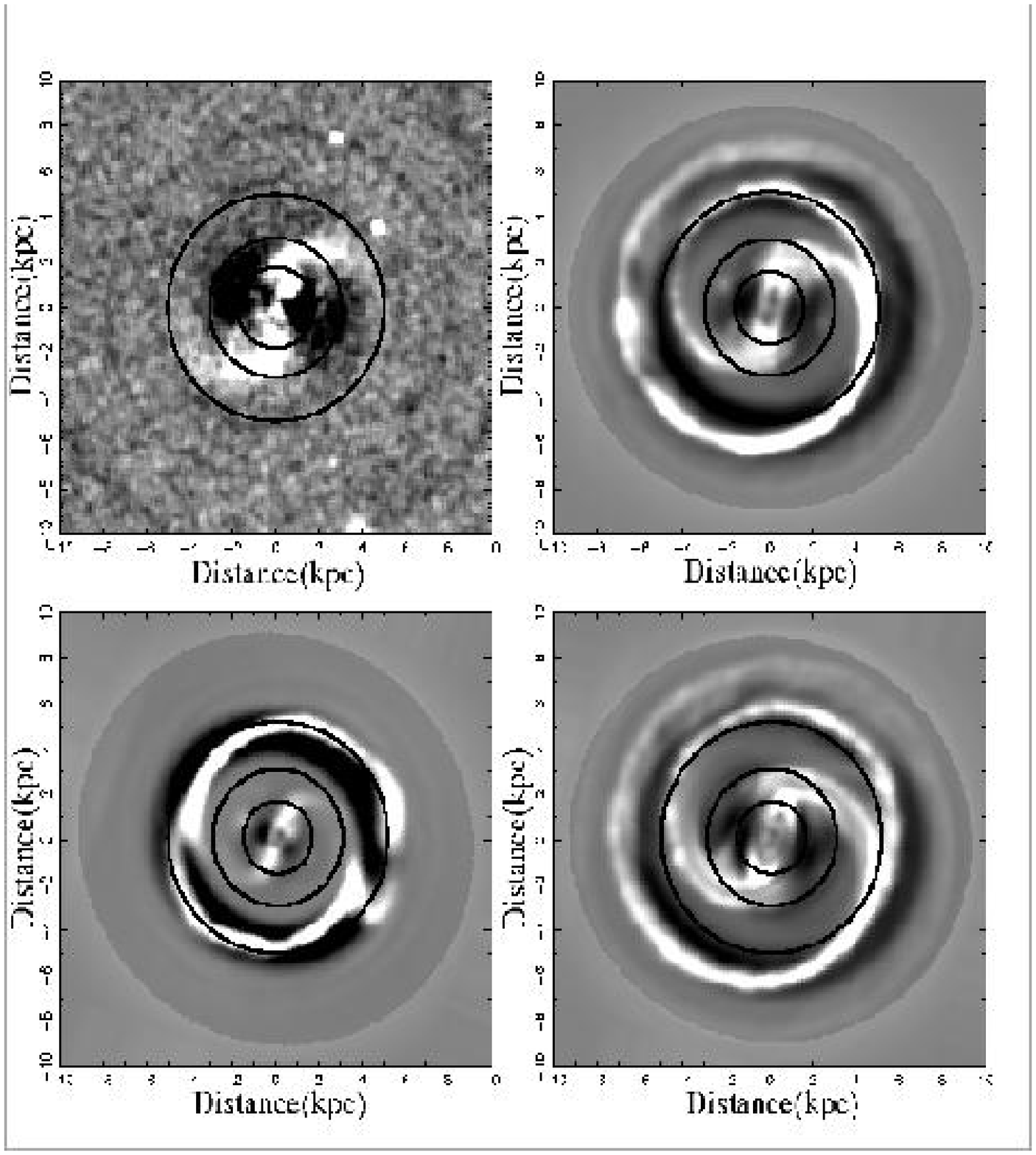,width=10.0cm}
\vspace{0cm}
\caption[Morphological comparison of IC~5186]{The  top left figure
shows the masked $B$-band image of IC~5186. The top right figure shows
the convolved and masked modelled gas density map for the IC~5186
model with $R_{\rm CR}$/$R_{\rm bar}$=1.4. The circles indicate the
position of the resonances for the axisymmetric approximation: from
inside to outside the IUHR, CR and OLR. The two lower figures show the
corresponding maps for the models with corotation at 1.1 and 1.6 times
the bar semi-major axis, from left to right respectively. In all
figures, the resonances are drawn for the $R_{\rm CR}$/$R_{\rm
bar}$=1.4 case as a reference.}
\label{fig:mask}
\end{center}
\end{figure*}

Although we cannot directly compare the modelled and observed gas
distributions, a comparison with the masked $B$-band image was carried
out as explained in Section~\ref{thebestfitmodels}. One can see from
Fig.~\ref{fig:mask} that the agreement in the inner regions is very
good for $R_{\rm CR}$/$R_{\rm bar}$~=~1.4, except that no sign of the
ring which formed in the simulation is observed on the $B$-band
image. It might be that there is no longer gas in the inner ring. High
spatial resolution imaging of the gas, maybe CO observations, would be
needed to disentangle the morphology of the gas in the inner parts and
compare it to the model gas density distribution.\\

When doing the same for decreasing pattern speeds, the main structural
features in the model move radially outward relative to the observed
structures. It seems that in order to match the main features
spatially some fine-tuning of the pattern speed would be required,
giving a slightly slower bar pattern speed.

\subsection {Model with mass distribution derived from the $I$-band light
distribution}
\label{modelwithmassdistributionderivedfromtheibandlightdistribution}

Models were run for a mass distribution derived from the $I$-band
luminosities to test the effect on the resulting gas distribution and
gas kinematics of using a photometric band more affected by
extinction. These models were run for three different pattern speeds,
and apart from the mass distribution the other initial conditions were
identical to those used to simulate the composite \HI-band image of
IC~5186. The light distribution for the $H$-band is smoother than that
of the $I$-band, which shows a more clumpy structure along the bar
similar to that observed in the $B$-band (Fig.~\ref{fig:light}). One
can also see the effect of dust in the fact that the inner ring is
brighter on one side of the galaxy.\\

\begin{figure}
\begin{center}
\centerline{\vbox{ \psfig{figure=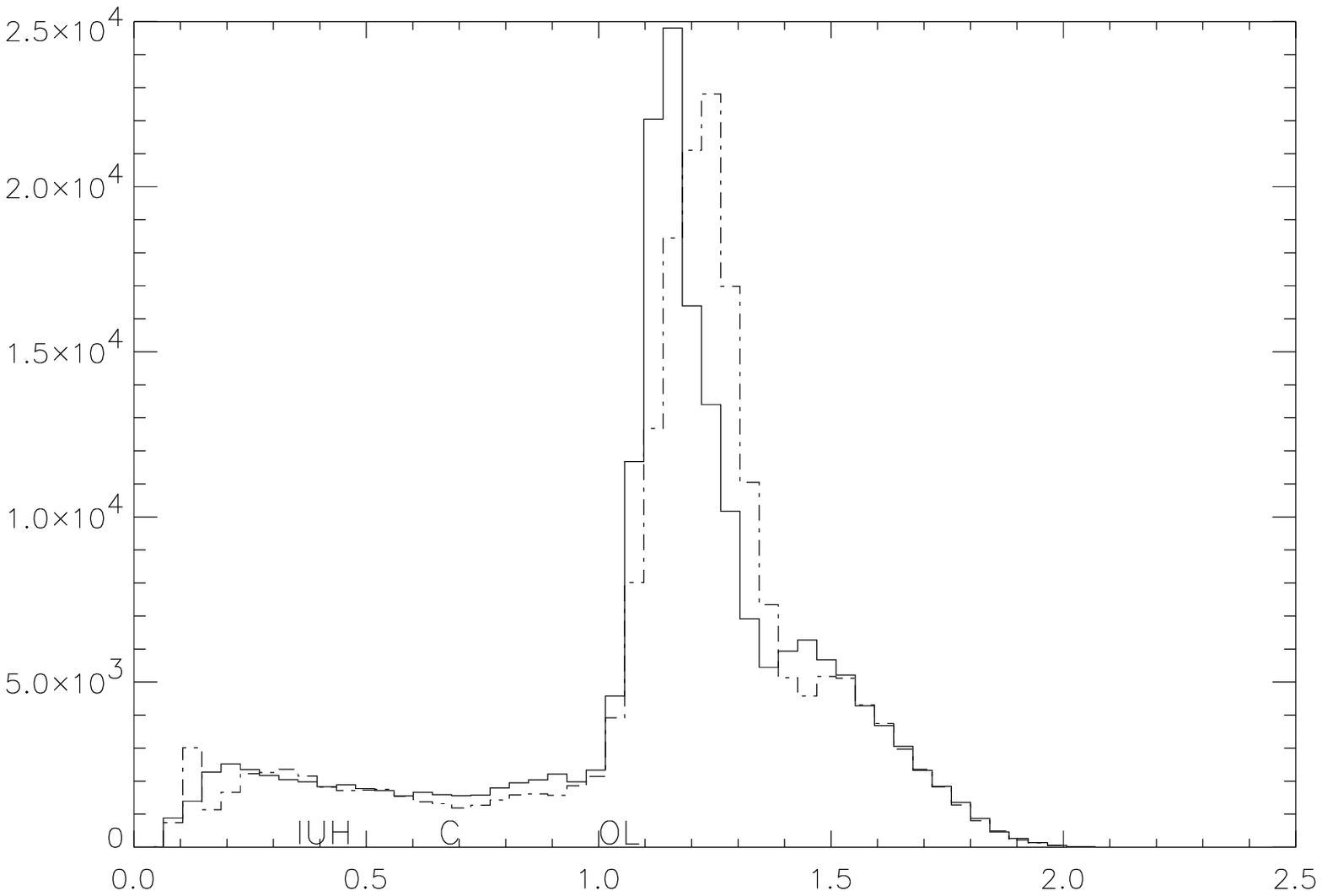,width=9.0cm} }}
\caption[Histogram of number of particles against angular momentum for
IC~5186]{Histogram of the number of particles  against angular
momentum for IC~5186 for the composite \HI-band image (solid line) and
the $I$-band image (dot-dash line) models. The distributions are shown
after 7 bar rotations. Most of the gas particles are located beyond
after the outer Lindblad resonance (OLR). The nuclear ring in the
$I$-band model is clearly visible as the small peak in number of
particles close to $Lz$=0.1}
\label{fig:Lz_IC5186}
\end{center}
\end{figure}

\begin{figure*}
\begin{center}
\vspace{0cm} \hbox{\epsfxsize=7.0cm
\epsfbox{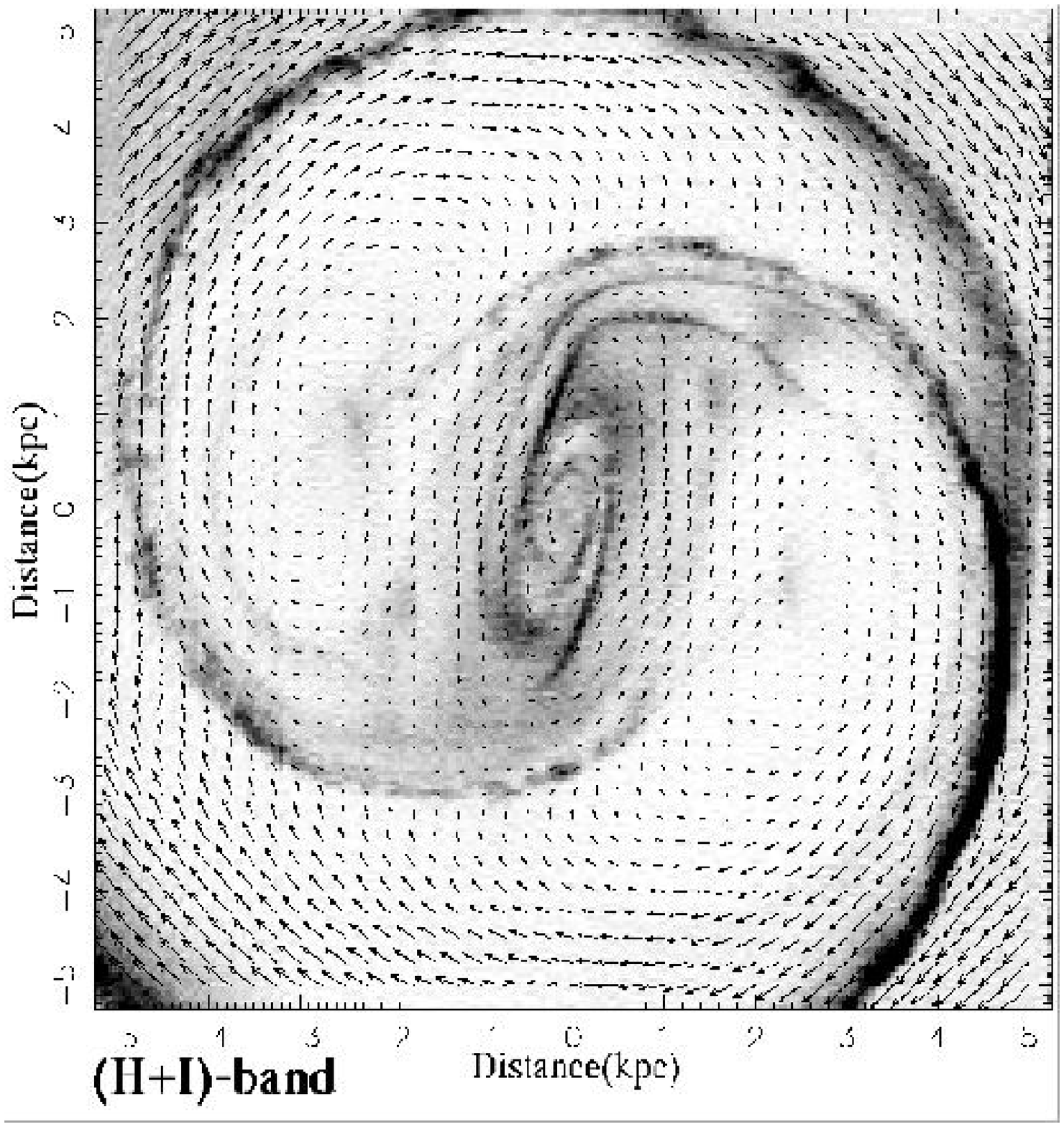} \epsfxsize=7.0cm
\epsfbox{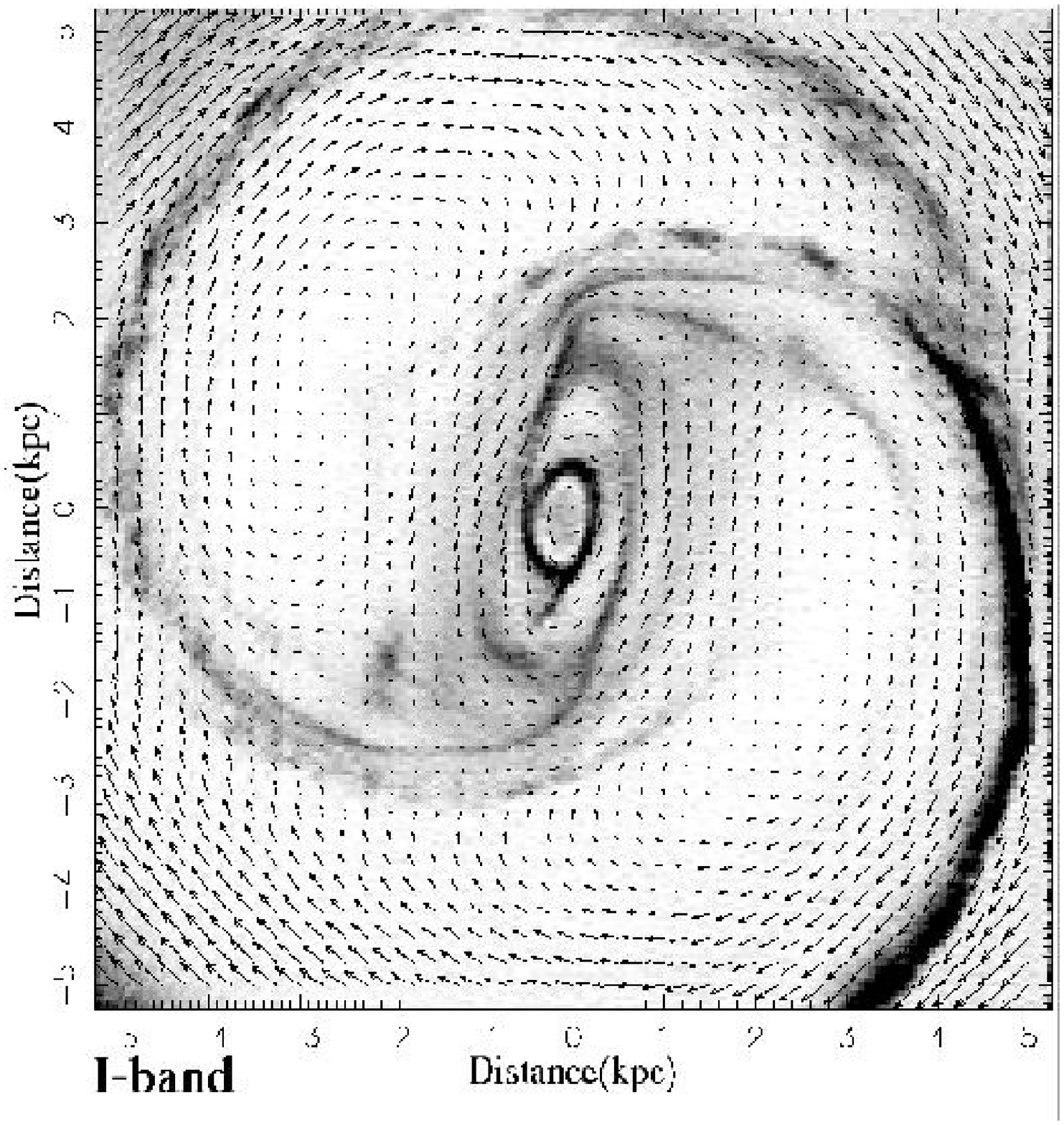}}
\vspace{0cm}
\caption[The gas distribution and velocity field for the mass
distributions of IC~5186: Comparison between the models derived from
the $I$-band image and those derived from the composite image]{The gas
distribution and velocity field for the mass distributions of IC~5186
derived  from the composite \HI-band image distribution (right) and
the $I$-band (left panel), after 7 bar rotations and for $R_{\rm
CR}$/$R_{\rm bar}$=1.4. The length of the vectors is proportional to
the velocity in the frame rotating with the bar. Notice the clear
inner ring which develops in the $I$-band simulations. The $H$-band
simulation does not show this ring so clearly, however part of the gas
particles have been trapped around the L$_{4}$ and the L$_{5}$
Lagrangian points (which are the zero velocity locations near the
direction perpendicular to the bar) this trapping also occurs for the
$I$-band simulation but to a lesser degree. }
\label{fig:HvsI}
\end{center}
\end{figure*}

The modelled gas distribution shows a striking difference. The nuclear
ring that is barely visible in the models derived from the \HI-band
image is the main morphological feature in the gas distribution of the
$I$-band model, see Fig.~\ref{fig:HvsI}. In order to determine the
nature of the nuclear ring the angular momentum distribution for both
cases was examined (see Fig.~\ref{fig:Lz_IC5186}). The main peak
observed for both distributions corresponds to the outer ring. Most of
the particles lie outside the outer Lindblad resonance (OLR), simply
reflecting the extended initial gas distribution. The low angular
momentum peak for the $I$-band case corresponds to the nuclear ring
seen in this case but barely seen on the composite image run. For the
$I$-band run there seems to be movement of particles from inside
corotation to populate the innermost ring.\\

The inner ring seems to be related to the inner ultra harmonic
resonance (IUHR), showing a four cornered structure. The nuclear ring
does not seem to be associated with any resonance and it is more
likely linked to the $x1$ family of orbits in the bar. The $H$-band
models seem to have trapped more gas around the $L_{4}$ and the
$L_{5}$ Lagrangian points. Fig.~\ref{fig:HvsI} suggests that the
nuclear ring more clearly present in the $I$-band case is made out of
the particles that are trapped around the $L_{4}$ and the $L_{5}$
Lagrangian points and in the inner ring in the \HI-band case. The
density enhancement near the Lagrangian points $L_{4,5}$ is displaced
azimuthally relative to $L_{4,5}$ in both cases (see
Fig.~\ref{fig:HvsI}). Stability of the Lagrangian points could be
checked to establish whether the density enhancement around the $L_{4}$
and the $L_{5}$ point is due to the fact that these points might be
stable. However, the fact that the potential is not bi-symmetric may
complicate the derivation of the Lagrangian points. This question will
be further addressed in a subsequent orbit analysis paper (in prep.).
The inner ring seems to have shrunk further in the \HI-band model
than in the $I$-band case, reflecting a higher degree of shock
dissipation. In the axisymmetric approximation there is no ILR at this
pattern speed in both cases. However, to identify properly the
resonances a detailed orbit analysis is required. The
position-velocity-diagram derived from $I$-band and from the composite
image do not show significant differences. In order to conclude
something about which model gives the best representation of the
galaxy dynamics 2-D kinematic information would be needed.\\

Summarising, the models derived for the mass distribution from the
$I$-band alone and from the $H$ and the $I$ band seem to give similar
overall results. There is a difference in the inner parts, where the
$I$-band derived model develops a clear nuclear ring whereas the
(\HI)-band derived model develops the nuclear ring but it is not so
marked. However, in this case there is a larger number of particles
populating the region around the $L_{4,5}$ Lagrangian points. There
are differences in the light distribution in the inner 3 kpc
region. The mass model derived from the $I$-band image presents a more
massive central core compared to the \HI-band mass distribution. Since
the $H$-band is a better tracer of the stellar mass one might expect
the opposite. However, as discussed in paper II, the M/L$_{\rm I}$ is
more sensitive to galaxy color changes than the M/L$_{\rm H}$. The
central regions of IC~5186 are more affected by extinction in the
$I$-band and therefore redder (though also dimmer) resulting in a
over-estimation of the M/L$_{\rm I}$. Increasing the central mass
concentration would favours the nuclear ring formation, as observed in
simulations by different authors \citep{regan}.

\subsection{Comparison of the models for different scale-heights}
\label{comparisonofthemodelsfordifferentscaleheights}
\begin{figure*}
\begin{center}
\vspace{0cm} \hbox{\hspace{0cm}\epsfxsize=7.0cm
\epsfbox{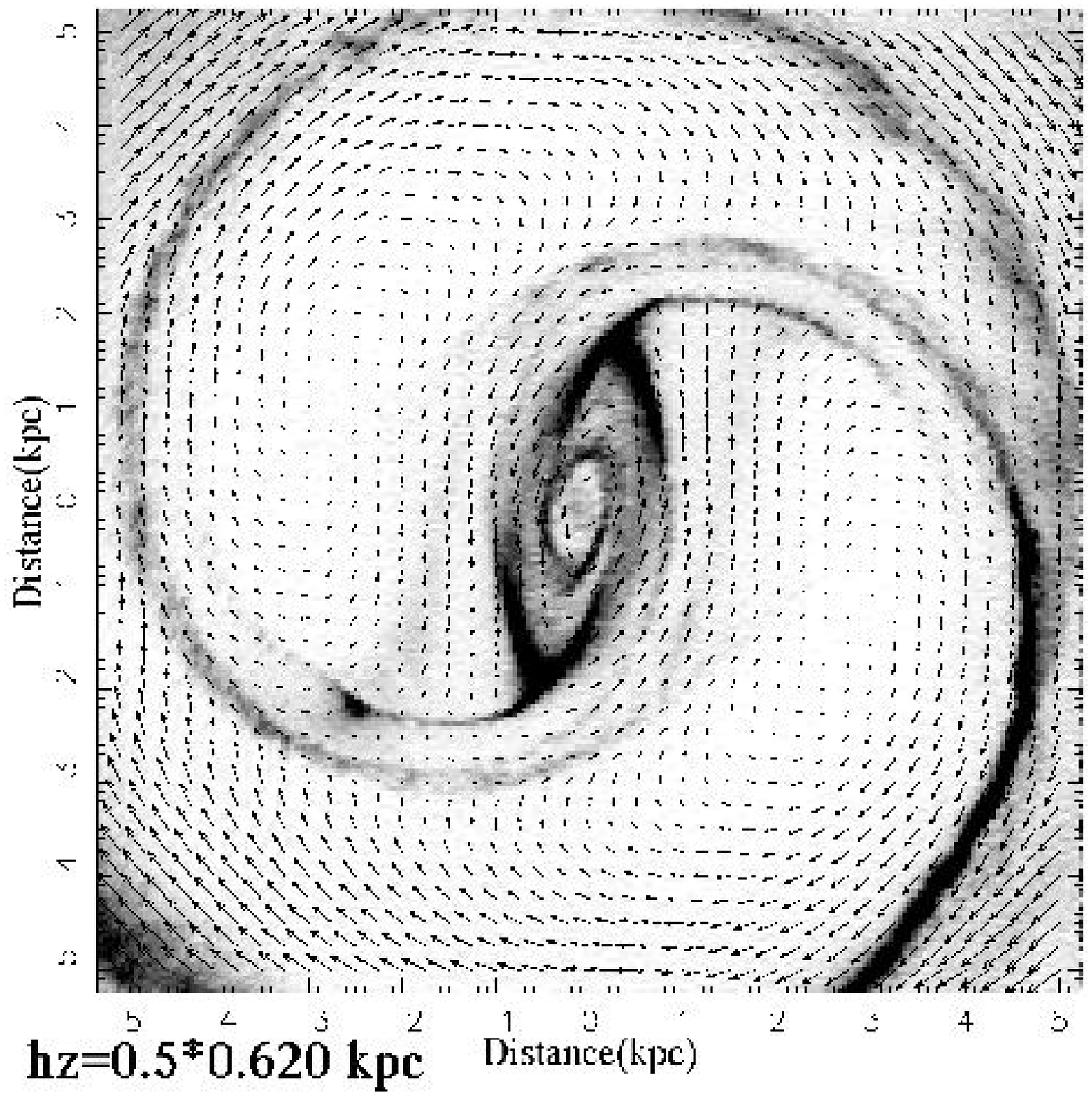} \epsfxsize=7.0cm \epsfbox{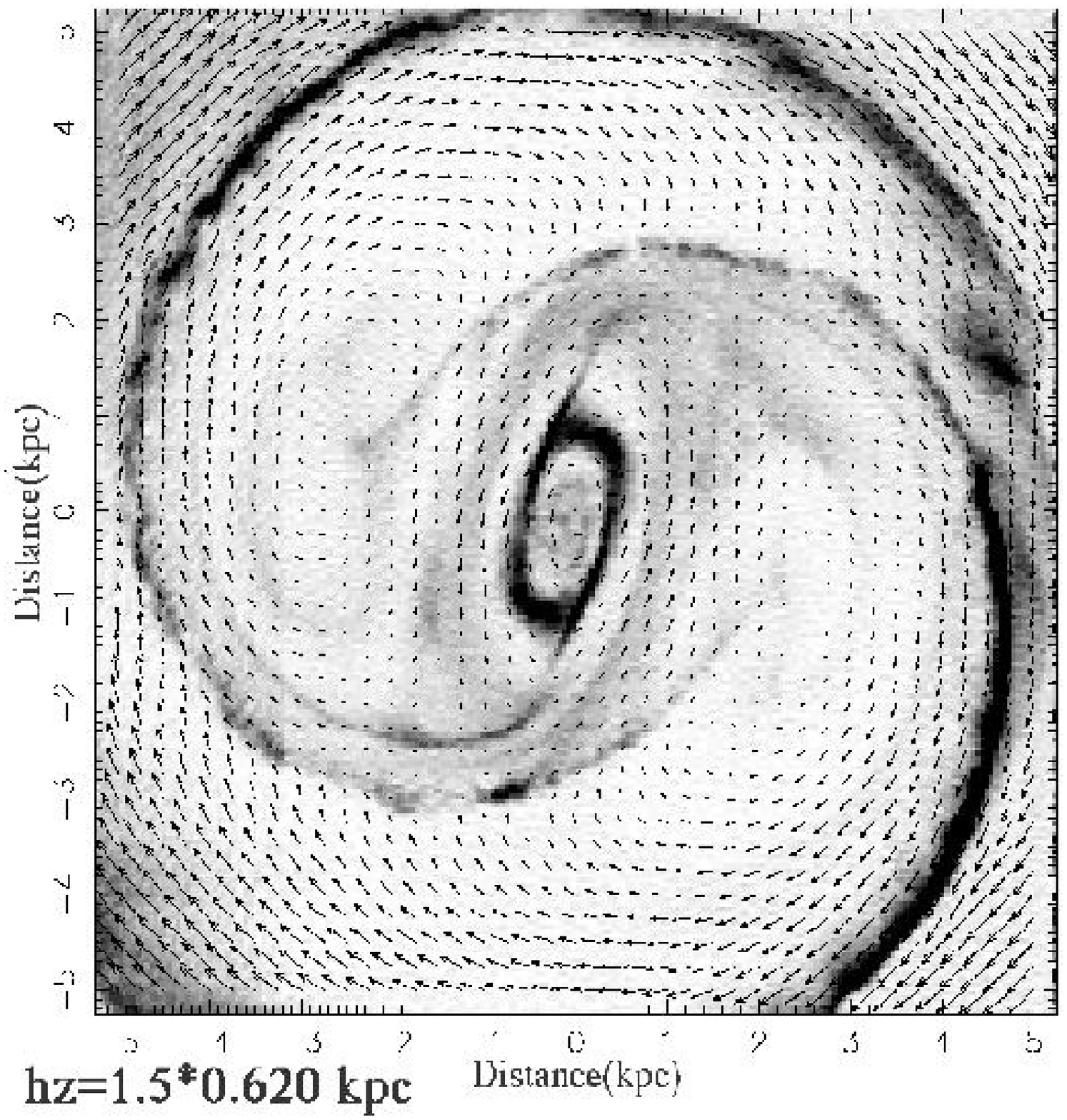}}
\vspace{0cm}
\caption[Velocity field of IC~5186 with different scale-heights]{ The
gas distribution and velocity field for the mass distribution of IC
5186 derived from the composite \HI-band image and with a scale-height
of 0.5 (left panel) and 1.5 (right panel) times the adopted standard
one, after 7 bar rotations and for $R_{\rm CR}$/$R_{\rm bar}$=1.4. The
length of the vectors is proportional to the velocity in the frame
rotating with the bar. The gas distributions show very different
features. There seems to be more gas near the lateral Lagrangian
points for a larger scale height; in fact the amount of gas there
seems to be related to the softening of the potential, either via
thickening the disk or via axisymmetrising it with a dark halo
component, as illustrated in  Fig.~\ref{fig:dm}}
\label{fig:heights}
\end{center}
\end{figure*}

\begin{figure*}
\begin{center}
\centerline{\vbox{ \psfig{figure=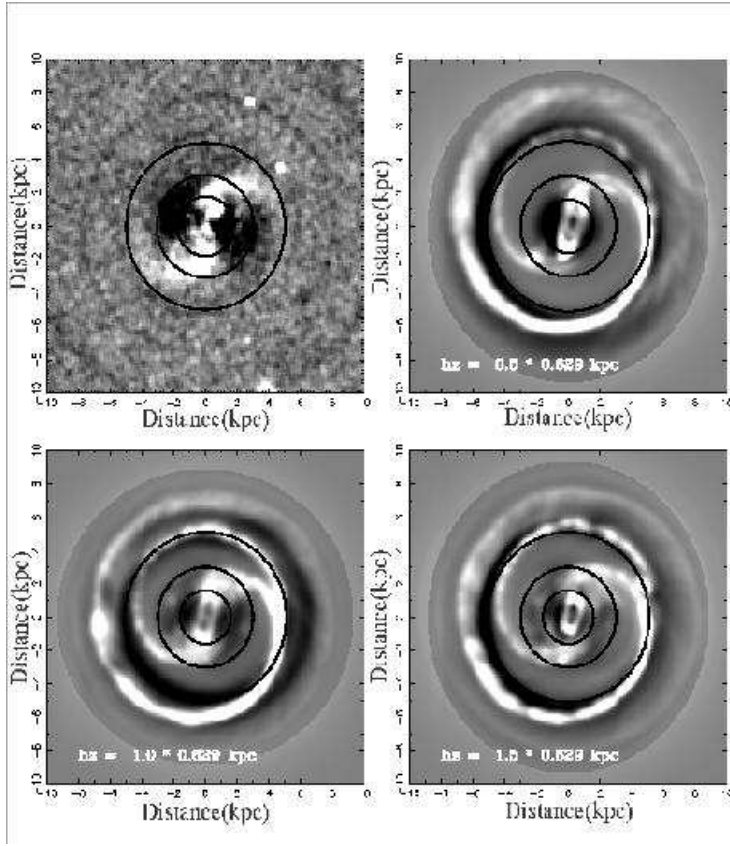,width=10cm} }}
\caption[Morphological comparison of IC~5186 using different
scale-heights]{The top left figure shows the masked  $B$-band image of
IC~5186. The lower left figure shows the convolved and masked modelled
gas density map of IC~5186 with $R_{\rm CR}$/$R_{\rm bar}$=1.4 and the
standard scale-height. The circles indicate the position of the
resonances in the axisymmetric approximation: from inside to outside,
the IUHR, CR and OLR. The other figures show the corresponding maps
for the models with scale-height of 0.5 (top right) and 1.5 (bottom)
right) times the standard one and the same pattern speed.  In all
figure, the resonances are drawn for the standard scale-height case a
a reference. The $B$-band map seems to be better represented by the
model with the standard scale-height.}
\label{fig:mask_height}
\end{center}
\end{figure*}

\begin{figure*}
\begin{center}
\centerline{\vbox{ \psfig{figure=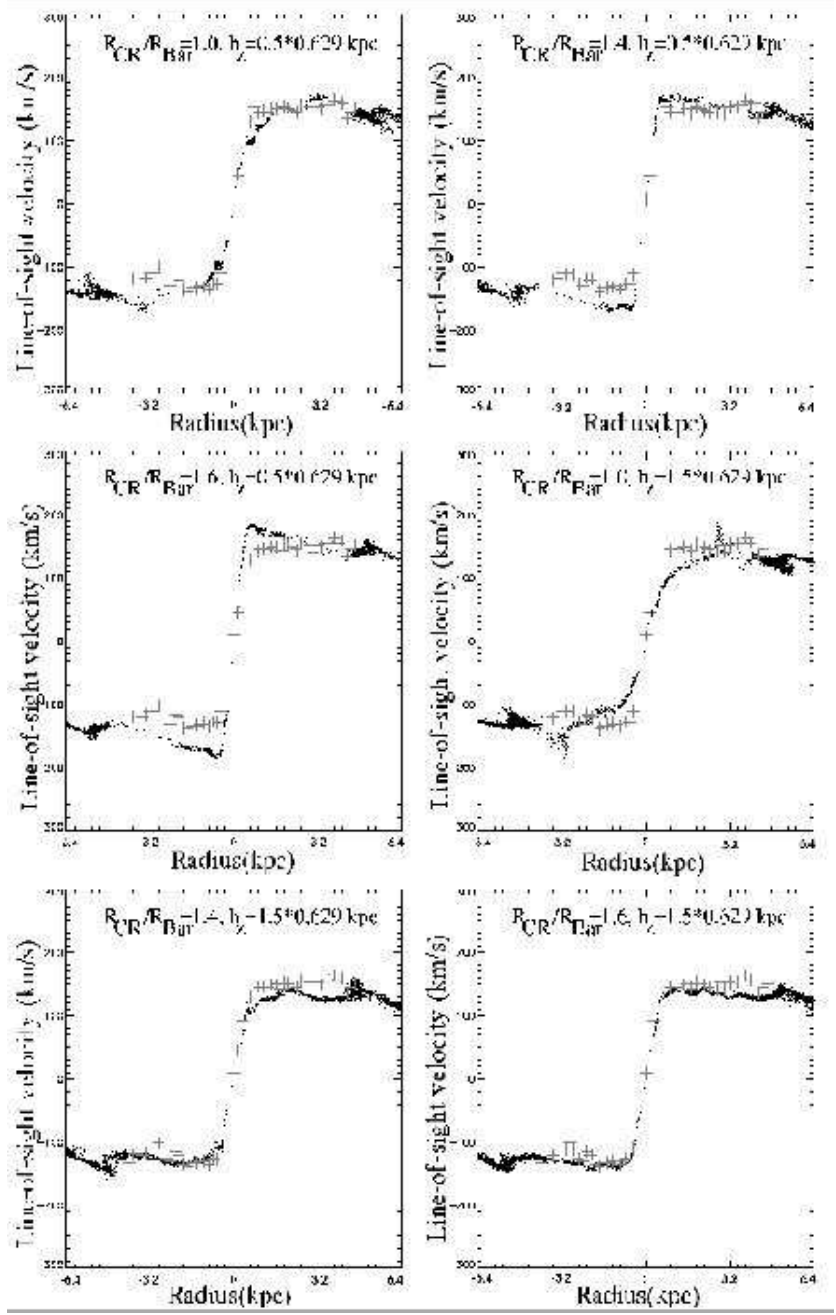,width=11cm} }}
\caption[Position-velocity diagrams along the major axis for IC~5186
for different scale-heights]{Line-of-sight velocity curve along the
major axis for the IC 5186 models with different bar pattern and
different scale-heights, after 8 bar rotations. The dots represent the
gas particles and the overlaid gray crosses the observed major axis
line-of-sight velocity curve. The $R_{\rm CR}$/$R_{\rm bar}$ and
scale-height for each position-velocity diagram is indicated in the
panels. The models do not agree with the data for the short
scale-height. Although the fit is not bad for the larger scale-height
at large $R_{\rm CR}$/$R_{\rm bar}$, such value for the scale-height,
given the size of the galaxy, is not observed in real galaxies.}
\label{fig:height_rc}
\end{center}
\end{figure*}

The goal of this test is to check the effect of changing the
scale-height in the modelled gas density distribution and the modelled
position-velocity diagrams. In addition to the simulations with the
adopted standard scale-height $h_z$, simulations with two other
scale-heights were run, i.e.  with 0.5 $\times$ $h_{\rm z}$ (0.31 kpc)
and 1.5 $\times$ $h_{\rm z}$ (0.94 kpc).  As one can see from
Fig.~\ref{fig:heights} a change in scale-height produces a significant
change in the gas distribution in the inner parts. From the velocity
field one can see that the shocks for the models with lower
scale-height are stronger as one would expect since these models are
enhancing the non-axisymmetry in the potential. The larger
scale-height produces less shocks since the increase in the
scale-height acts as a softening. Fig.~\ref{fig:mask_height} shows the
comparison between the masked observed $B$-band and the masked models
with different stellar scale-heights. The modelled position-velocity
diagrams (see Fig.~\ref{fig:height_rc}) show a large difference with
respect to the observed rotation curve for the short scale-height and
none of the pattern speeds give good fit; for the models with 1.5
$\times$ $h_{\rm z}$ the best fit corresponds to the pattern speed
with corotation at 1.6 times the bar semi-major axis. However, a scale
height of  1.5$\times$ $h_{\rm z}$ (0.94 kpc) is an unrealistically
large value for the vertical scale of a real galaxy with the observed
scale-length \citep{kregel02}.

\section{NGC 5728}
\label{ngc5728}

NGC 5728 is SABa Seyfert type 2 galaxy at a distance of 35 Mpc. It
shows an outer and inner stellar ring and there is also evidence for
the presence of a secondary stellar bar as reported by \citet{wozniak95}. There is no nuclear bar structure in the
molecular gas and at larger scales the CO emission traces the primary
bar and the outer ring structure \citep{combes02}. The bar
position angle is $\approx$ 30$^{\circ}$ to the line of nodes.\\

\begin{figure}
\begin{center}
\vspace{0cm} \hbox{\hspace{0cm}\epsfxsize=4.5cm
\epsfbox{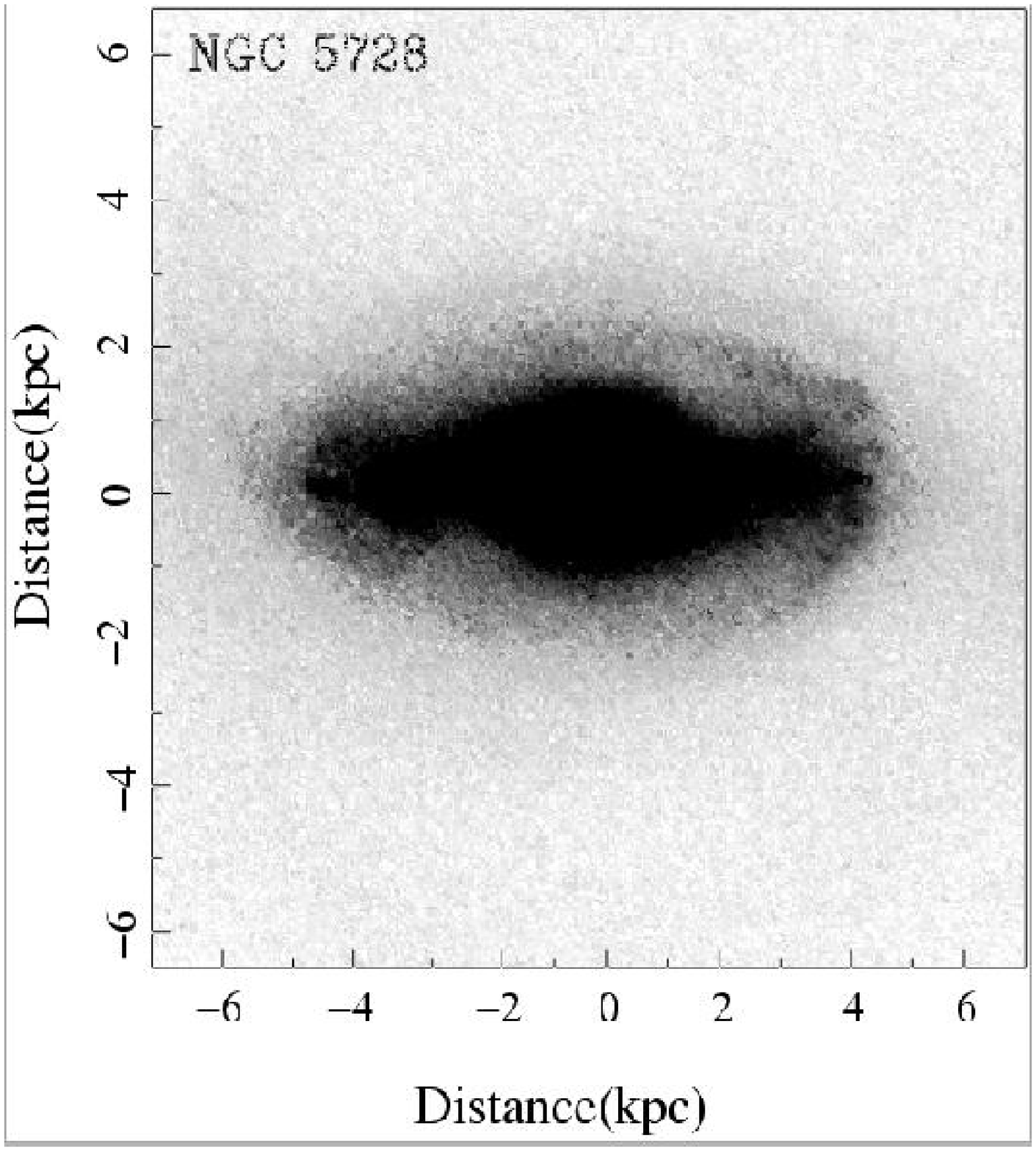} \epsfxsize=4.5cm
\epsfbox{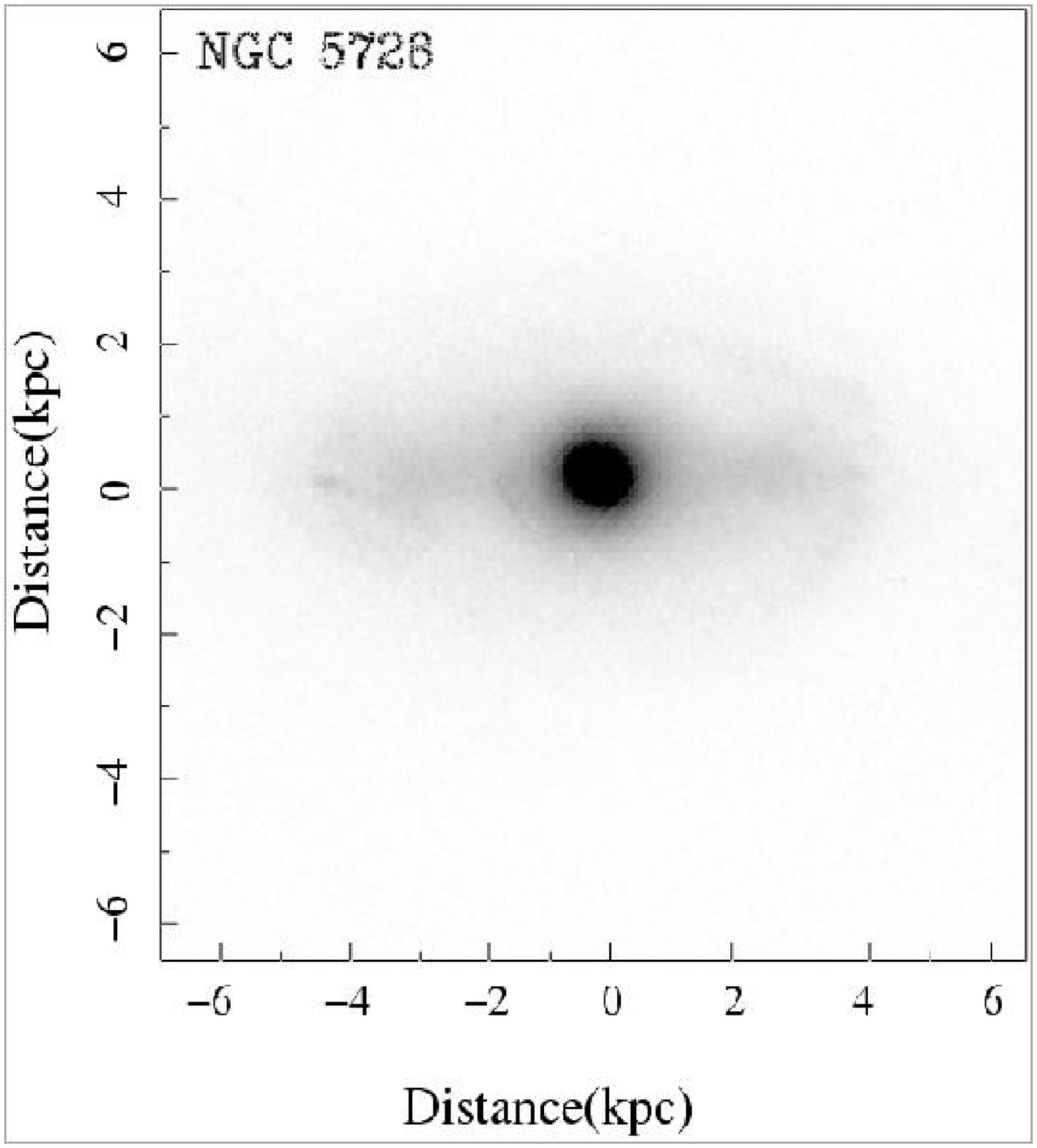}}
\vspace{0cm}
\caption[Light distribution for NGC 5728]{Deprojected light
distribution in the $H$-band of the inner part of NGC~5728. The left
panel shows the extent of the primary bar and the right panel the
extent of the nuclear structure.}
\label{fig:barra}
\end{center}
\end{figure}

\begin{figure*}
\begin{center}
\vspace{0cm}
\hspace{0cm}\psfig{figure=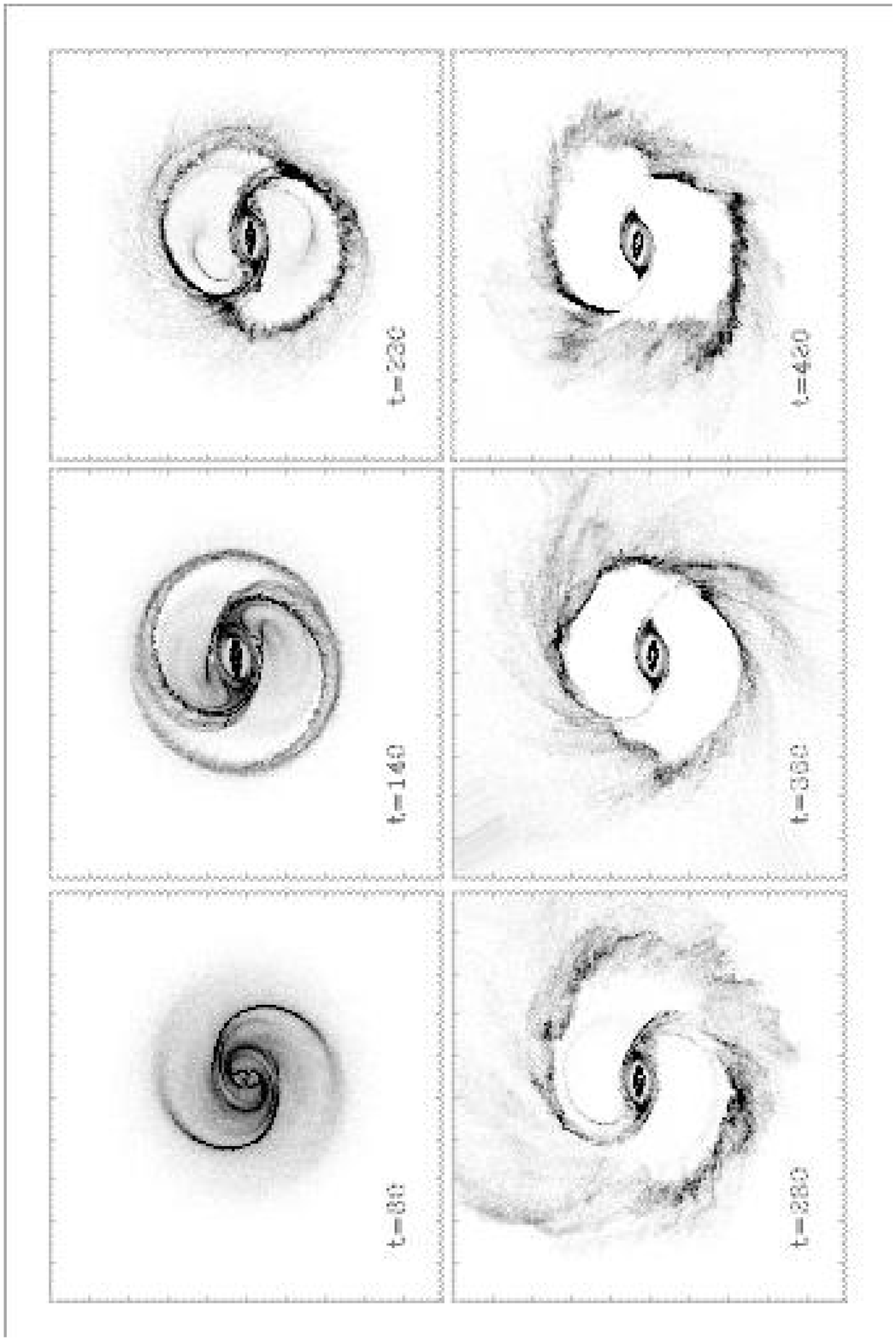,width=10.0cm,angle=-90}
\vspace{0cm}
\caption[Time evolution of the gas distribution for NGC5728
simulation]{Time evolution of the gas distribution in the NGC5728
simulation  with corotation at the end of the bar
($R_{CR}/R_{bar}$=1.0), the bar pattern is rotating
counter-clockwise. Each panel, from left to right and top to bottom,
shows the distribution at different integration times, from 1 bar
rotation to 6 bar rotation, and in the frame corotating with the
bar. The non-axisymmetric component is fully grown after three bar
rotations and the bar pattern is rotating counter-clockwise in the
inertial frame. The size of the frames is 50 kpc in diameter and time
is in Myr.}
\label{fig:ngc5728evolgas}
\end{center}
\end{figure*}

\begin{figure}
\begin{center}
\vspace{0cm}
\hspace{0cm}\psfig{figure=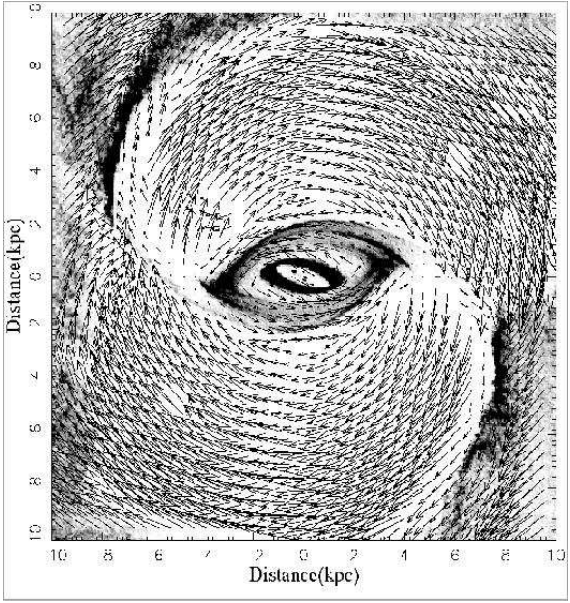,width=7.5cm}
\vspace{0cm}
\caption[Velocity field for NGC 5728]{The gas distribution and
velocity field for the mass distribution derived from the composite
\HI-band image of NGC~5728, at 6 bar rotations and for $R_{\rm
CR}$/$R_{\rm bar}$=1.0.  The length of the vectors is proportional to
the velocity in the  frame rotating with the bar.}
\label{fig:vfield_NGC5728}
\end{center}
\end{figure}

The models are integrated for six bar rotations (see Fig.~\ref{fig:ngc5728evolgas} with
$R_{\rm CR}$/$R_{\rm bar}$=1.0), Table~\ref{tab:gal_par} shows the
pattern speeds explored in the simulations. The spiral arms do not
remain stationary but since the simulations are not self-consistent
and we assume a single pattern speed this behaviour in the outer parts
is not surprising. A nuclear ring develops with an orientation
slightly tilted with respect to the inner ring
(Fig.~\ref{fig:vfield_NGC5728}) . The orientation of the nuclear ring
with respect to the inner ring does not vary in time. This nuclear
ring observed in the models has the same orientation as the secondary
bar in the real galaxy (see Fig.~\ref{fig:vfield_NGC5728} and
Fig.~\ref{fig:barra}). The inner ring lies very close to the
corotation zone with the gas following a clear high energy $x_{1}$
family orbital shape.
The frequency diagram in the axisymmetric approximation shows that for
$R_{\rm CR}$/$R_{\rm bar}$=1 two ILRs exist. However, the nuclear ring
is not oriented perpendicular to the large scale bar and thus does not
seem to be supported by $x_2$ orbits. If one compares the light
distribution in the $H$-band (see Fig.~\ref{fig:barra}) with the gas
density distribution (Fig.~\ref{fig:vfield_NGC5728}) one can observe
in the $H$-band image (Fig.~\ref{fig:barra}) that the gas distribution
seems to coincide with the observed stellar ring. The $H$-band image
suggests that the bar has the typical shape of $x_{1}$ orbits looping
near the apencenter. The gas cannot stay on such intersecting orbits,
therefore it is no surprise that no gas is found in the models on
these lower energy $x_{1}$ orbits. However, in order to identify
properly the various orbit families, an orbit analysis is necessary,
which is beyond the scope of this study, focused on the dark matter
content.\\

\begin{figure*}
\begin{center}
\vspace{0cm}
\hspace{0cm}\psfig{figure=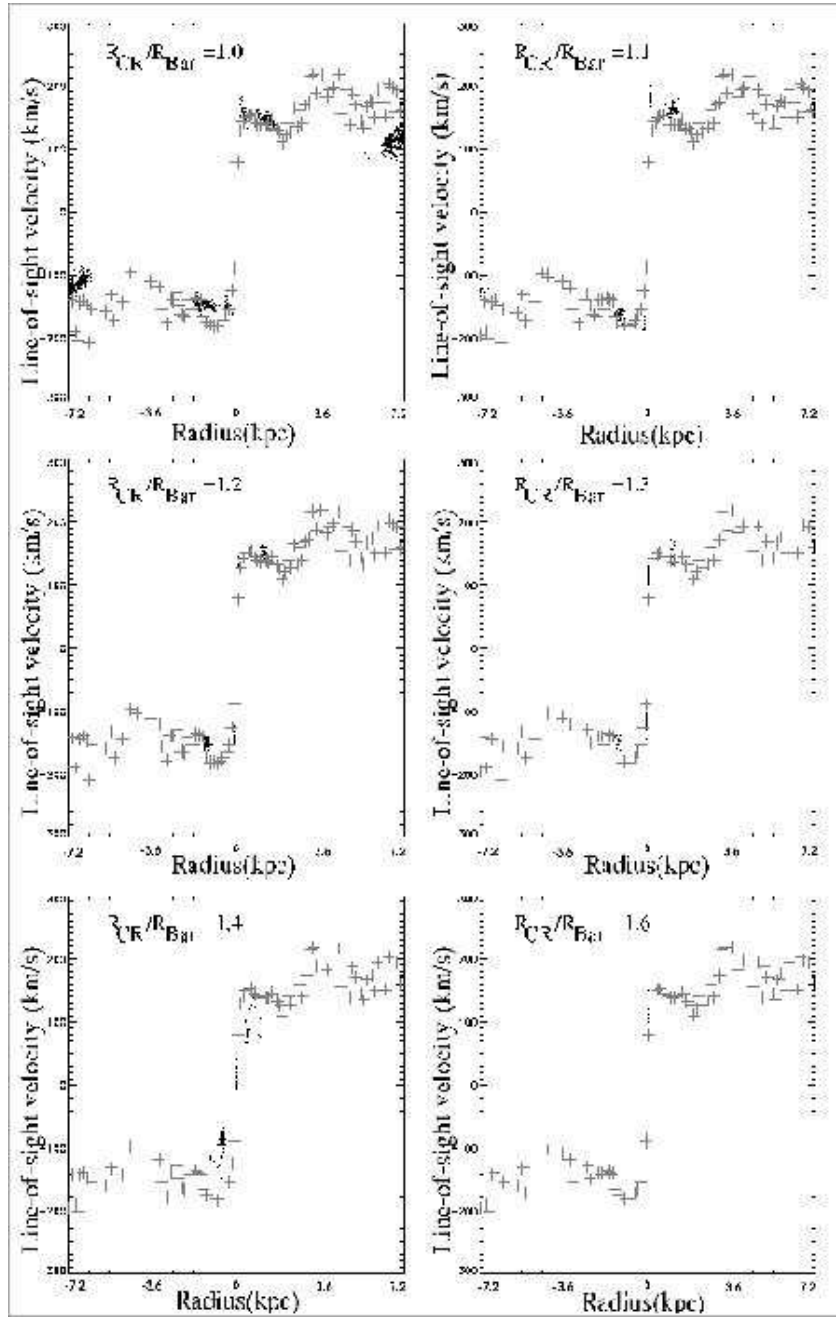,width=11.0cm}
\vspace{0cm}
\caption[Position-velocity diagrams along the major axis for NGC
5728]{Line-of-sight velocity curves along the major axis for the NGC~5728
models with different bar pattern speeds, after 6 bar rotations.  The
dots represent the gas particles and the overlaid gray crosses represent
the observed major axis line-of-sight velocity curve.}
\label{fig:rc_NGC5728}
\end{center}
\end{figure*}

There is not a very good agreement between the modelled major axis
line-of-sight velocity diagrams and the observed optical L-O-S velocity
curves for any pattern speed (see Fig.~\ref{fig:rc_NGC5728}), probably
due to the fact that in reality the inner bar rotates faster than the
primary bar. The torques due to this decoupled bar are lower than
those produced by an inner bar rotating at the same speed as the
primary bar, leading to a less efficient flow of gas toward the
centre \citep{combes94}. This would be supported by the strong
non-circular signature seen in the modelled velocity diagrams. One can also
notice in the velocity curves that the slower bars show a more centrally
concentrated gas distribution than the models with corotation closer
to the bar ends. The major axis L-O-S velocity curve with
$R_{\rm CR}$/$R_{\rm bar}$=1.0, 1.1 and 1.2 seem to give an overall
agreement of the rotation curve normalisation factor compared to the
observations, favouring the adopted M/L ratio calculated from
population synthesis models.

\section{NGC 7267}
\label{ngc7267}

This is a four armed barred SB(rs)a spiral at a distance of 44
Mpc. The bar is very bright and has complex structure in the
continuum, in effect splitting at the ends.
In H$\alpha$, the dominant feature traces the bar. But there are
bright complexes near the bar ends that do not correlate well with the
continuum features. No clear ring patterns are evident in H$\alpha$.
Discrete HII regions and diffuse emission lie around the bar and in
one of the outer arms \citep{crocker} .
There is an intense nucleus of emission. The bar position angle is
approximately at right angle with respect to the line of nodes.\\

\begin{figure*}
\begin{center}
\vspace{0cm}
\hspace{0cm}\psfig{figure=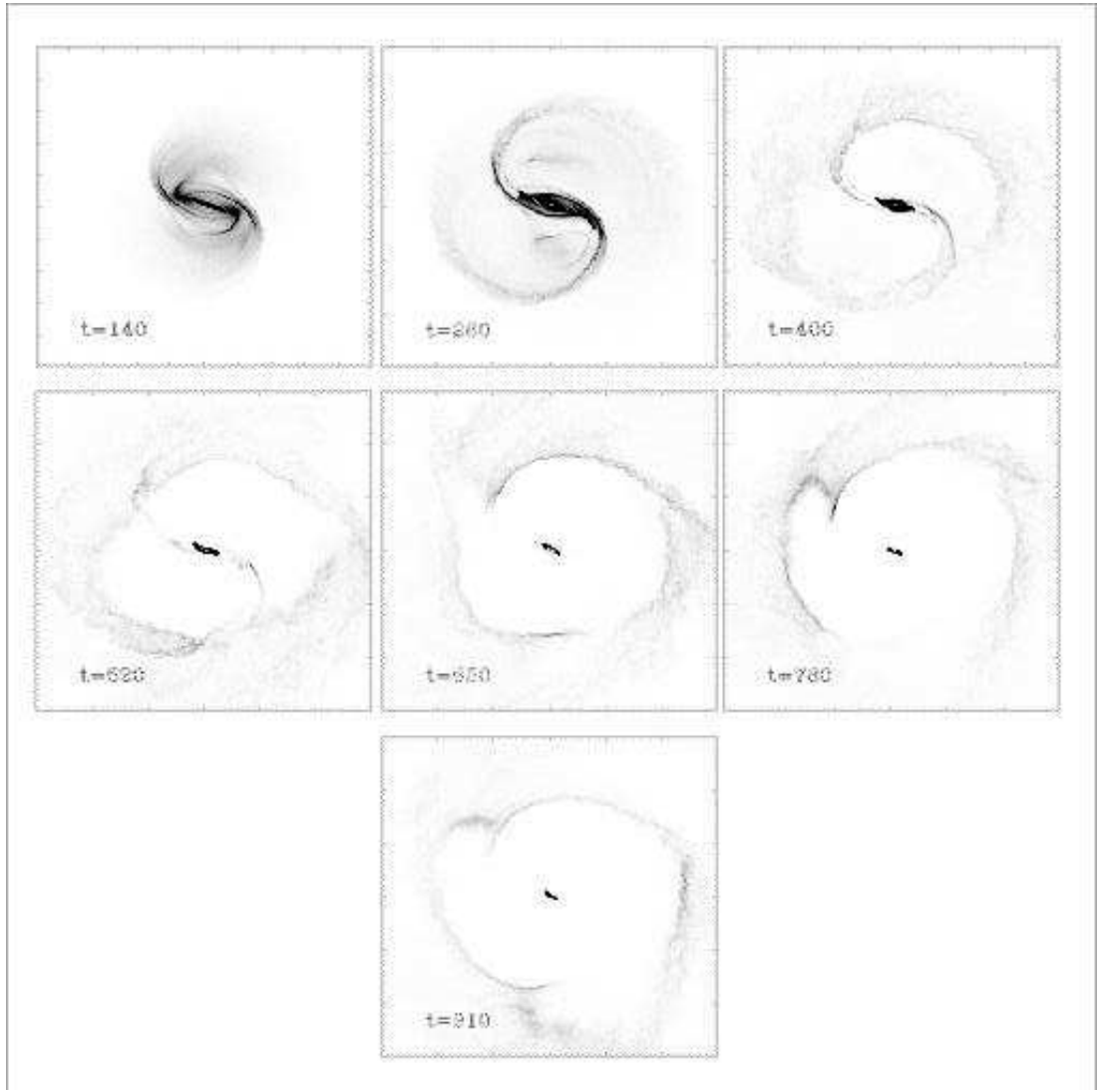,width=14.5cm,angle=-90}
\vspace{0cm}
\caption[Time evolution of the gas distribution in the NGC7267
simulation]{ Time evolution of the gas distribution in the NGC 7267
simulation with corotation at 1.0$\times$the bar semi-major axis. Each
panel, from left to right and top to bottom, shows the distribution at
different integration times, from 1 bar rotation to 7 bar rotations,
and in the frame corotating with the bar. The non-axisymmetric
component is fully grown after three bar rotations and the bar pattern
is rotating counter-clockwise in the inertial frame. The size of the
frames is 30 kpc in diameter  and time is in Myr. The gas does not
reach a steady state and the inner feature shrinks more and more after
every bar rotation.}
\label{fig:ngc7267evolgas}
\end{center}
\end{figure*}

The pattern speeds explored in the simulations are shown in
Table~\ref{tab:gal_par}. The simulations fail to reach a stationary
state for any of the pattern speeds
Fig.~\ref{fig:ngc7267evolgas}. Since no steady configuration was found
no comparison with the observed data was performed. A detailed
discussion of the possible reasons why the modelling fails are given
in Section ~\ref{discussionchapter5}. One a priori reason why the gas
fails to reach a steady state is the asymmetries in the light
distribution. To test this possibility the potential was calculated
from the bisymmetrised light distribution of NGC~7267 and the code was
run with the standard parameters. No steady state is reached either
for any of the pattern speeds.

\section{NGC 7483}
\label{ngc7483}

An outer and an inner ring are the main morphological features of the
SABa galaxy NGC 7483 at a distance of 66 Mpc. It also seems to posses
a nuclear ring perpendicular to the bar. There is no kinematic nor
photometric study of this galaxy in the literature. The bar position
angle is approximately at right angle to the line of nodes.\\
\subsection{Results}

\begin{figure}
\begin{center}
\vspace{0cm}
\hspace{0cm}\psfig{figure=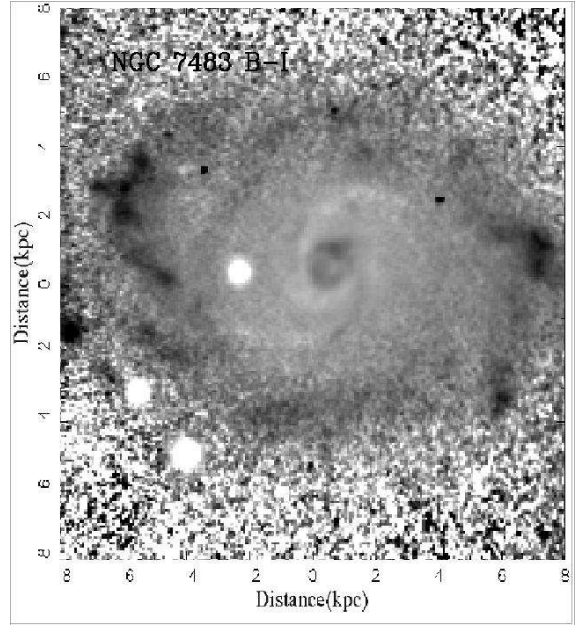,width=7.5cm}
\vspace{0cm}
\caption[$B$-$I$ map for NGC 7483]{$B$-$I$ map for NGC 7483}
\label{fig:bingc7483}
\end{center}
\end{figure}

\begin{figure*}
\begin{center}
\vspace{0cm}
\hspace{0cm}\psfig{figure=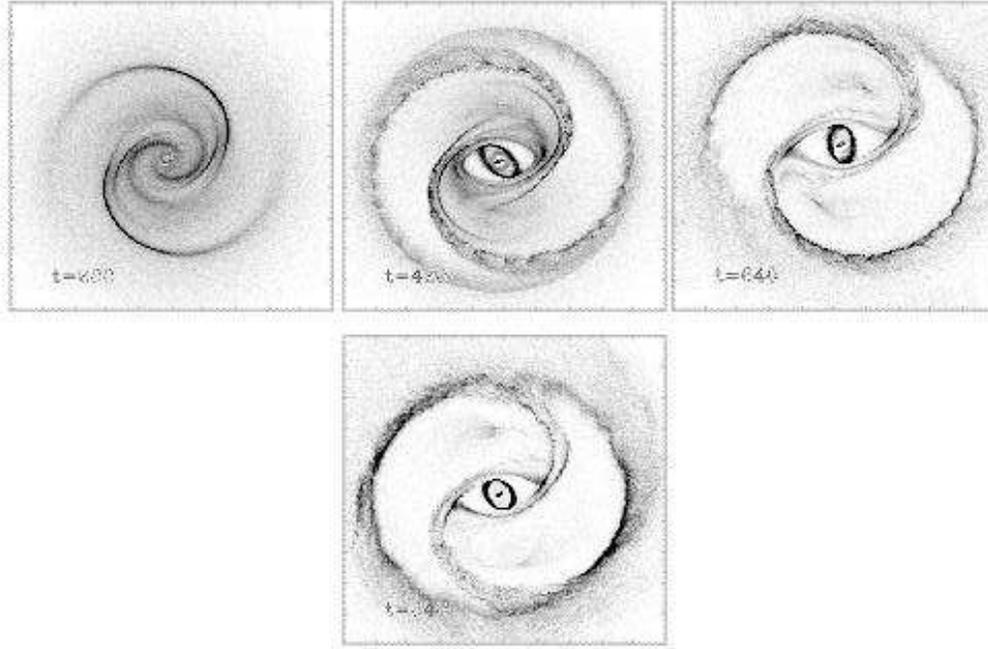,width=10.0cm,angle=-90}
\vspace{0cm}
\caption[Time evolution of the gas distribution in the NGC 7483
simulation]{ Time evolution of the gas distribution in the NGC 7483
simulation with corotation at 1.0$\times$the bar semi-major axis. The
bar pattern is rotating counter-clockwise. Each panel, from left to
right and top to bottom, shows the distribution at different
integration times, from 1 bar rotation to 4 bar rotations, and in the
frame corotating with the bar. The non-axisymmetric component is fully
grown after three bar rotations and the bar pattern is rotating
clockwise in the inertial frame. The size of the frames is 50 kpc in
diameter and time is in Myr.}
\label{fig:ngc7483evolgas}
\end{center}
\end{figure*}

\begin{figure}
\begin{center}
\vspace{0cm}
\hspace{0cm}\psfig{figure=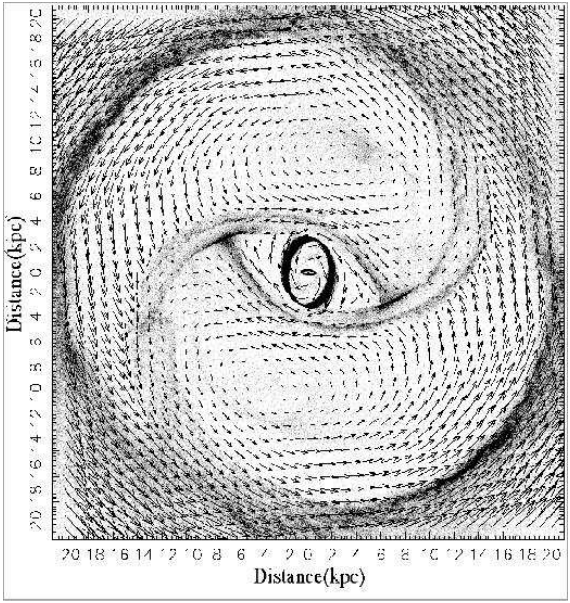,width=7.5cm}
\vspace{0cm}
\caption[Velocity field for NGC~7483]{The gas distribution and
velocity field for the mass distribution derived from the composite
\HI-band image of NGC~7483, at 4 bar rotations and for $R_{\rm
CR}$/$R_{\rm bar}$=1.0. The length of the vectors is proportional to
the velocity in the frame rotating with the bar.}
\label{fig:ngc7483_vfield}
\end{center}
\end{figure}

The pattern speeds explored in the simulation are presented in
Table~\ref{tab:gal_par}. The gas density distribution clearly
settles in a stationary configuration after three bar rotations and
for all pattern speeds a nuclear ring develops perpendicular to the
bar Fig.~\ref{fig:ngc7483evolgas}. From the frequency plot of the axisymmetric case this
galaxy possesses two ILRs for all the pattern speeds explored (in the
axisymmetric approximation).  Although the general gas configuration
remains stable the nuclear ring does not remain steady and wobbles in
time. Whether this ring is formed purely by particles confined to
$x_{2}$ orbits will have to be checked through an orbit
analysis. There is evidence from Fig.~\ref{fig:ngc7483_vfield} of large
velocity gradients and the velocity vectors have different directions
in the nuclear region indicating the presence of shocks. In fact, when
one looks at the $B$-$I$ map (see Fig.~\ref{fig:bingc7483}) the
presence of complicated dust lanes featuring several arm spiral
structures in the nuclear region is evident.

\begin{figure}
\begin{center}
\centerline{\vbox{ \psfig{figure=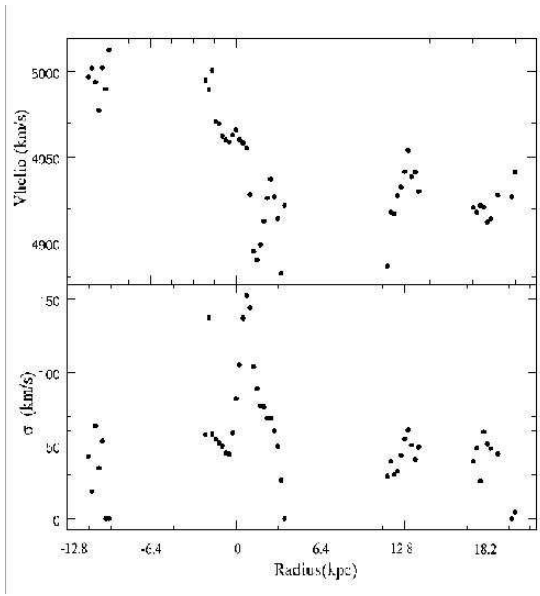,width=7.5cm} }}
\caption[Long slit spectroscopic observations of the $H_{\alpha}$ line
of NGC~7483]{Long slit spectroscopic observations of the $H_{\alpha}$
line of NGC 7483 with the slit oriented along the major axis. Notice
how messy the rotation curve appears and compare it to the modelled
rotation curve (Fig.~\ref{fig:pv_NGC7483})}
\label{fig:rc_NGC7483}
\end{center}
\end{figure}

\begin{figure*}
\begin{center}
\vspace{0cm}
\hspace{0cm}\psfig{figure=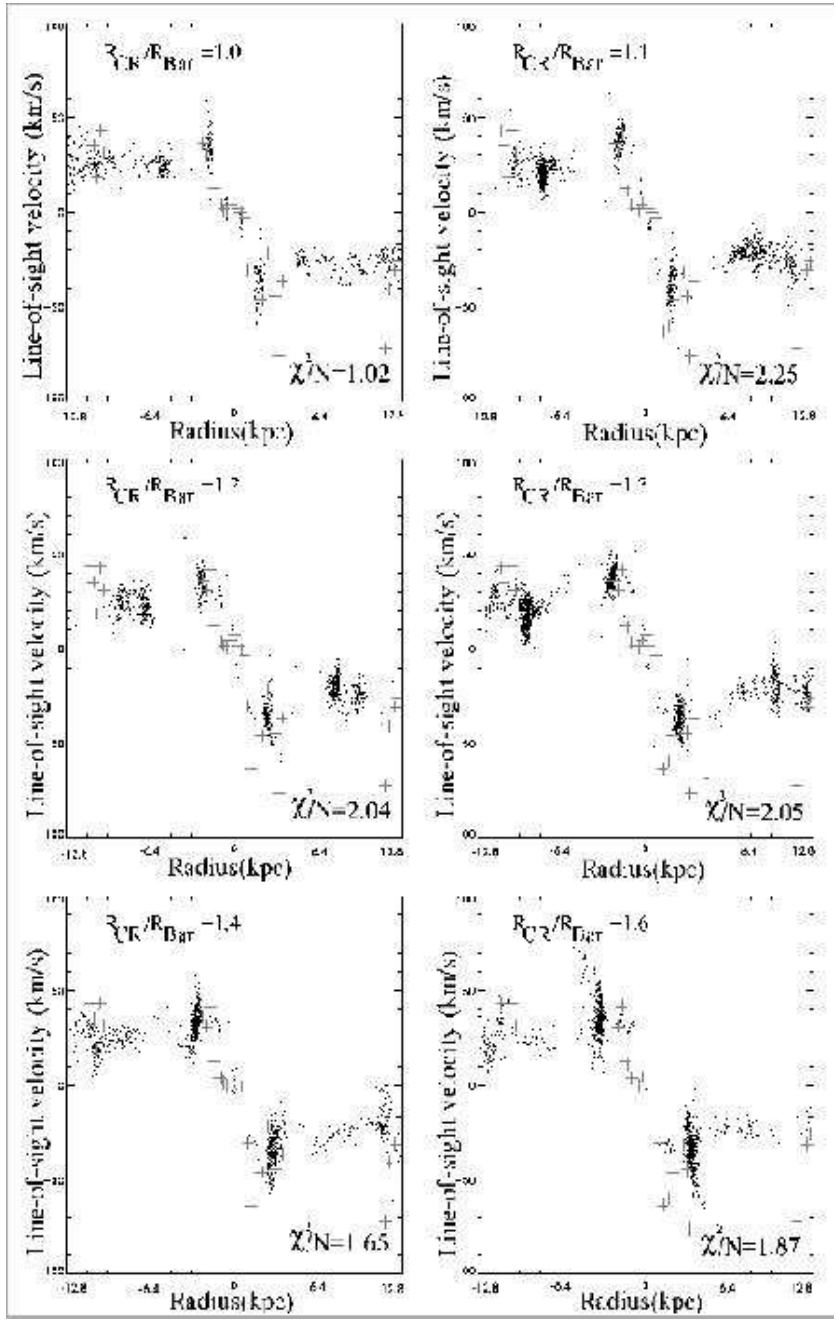,width=11.0cm}
\vspace{0cm}
\caption[Position-velocity diagrams along the major axis for
NGC~7483]{Line-of-sight velocity curves along the major axis for the
NGC 7483  models with different bar pattern speeds, after 4 bar
rotations. The dots represent the gas particles and the overlaid
gray crosses the observed major axis line-of-sight velocity curve. The
modelled velocity curve agrees very well with the observed
data. The best fit model is from the simulation with $R_{\rm
CR}$/$R_{\rm bar}$=1.0 (see Table~\ref{tab:gal_par})}
\label{fig:pv_NGC7483}
\end{center}
\end{figure*}

The kinematics obtained through long slit spectroscopy at the 1.5 m at
La Silla Observatory show a very complicated  rotation curve (see
Fig.~\ref{fig:rc_NGC7483}); however a comparison of the modelled
L-O-S velocity curve with the data shows a good agreement (see
Fig.~\ref{fig:pv_NGC7483}). As in the case of IC~5186 the amplitude of
the rotation curve, fixed by the choice of M/L, agrees well with the
observed rotation curve. The best fit is for a pattern speed  of 27.50
km s$^{-1}$ kpc$^{-1}$ corresponding to a corotation radius at the end
of the bar (see Table~\ref{tab:gal_par}).\\

\begin{figure*}
\begin{center}
\centerline{\vbox{
\psfig{figure=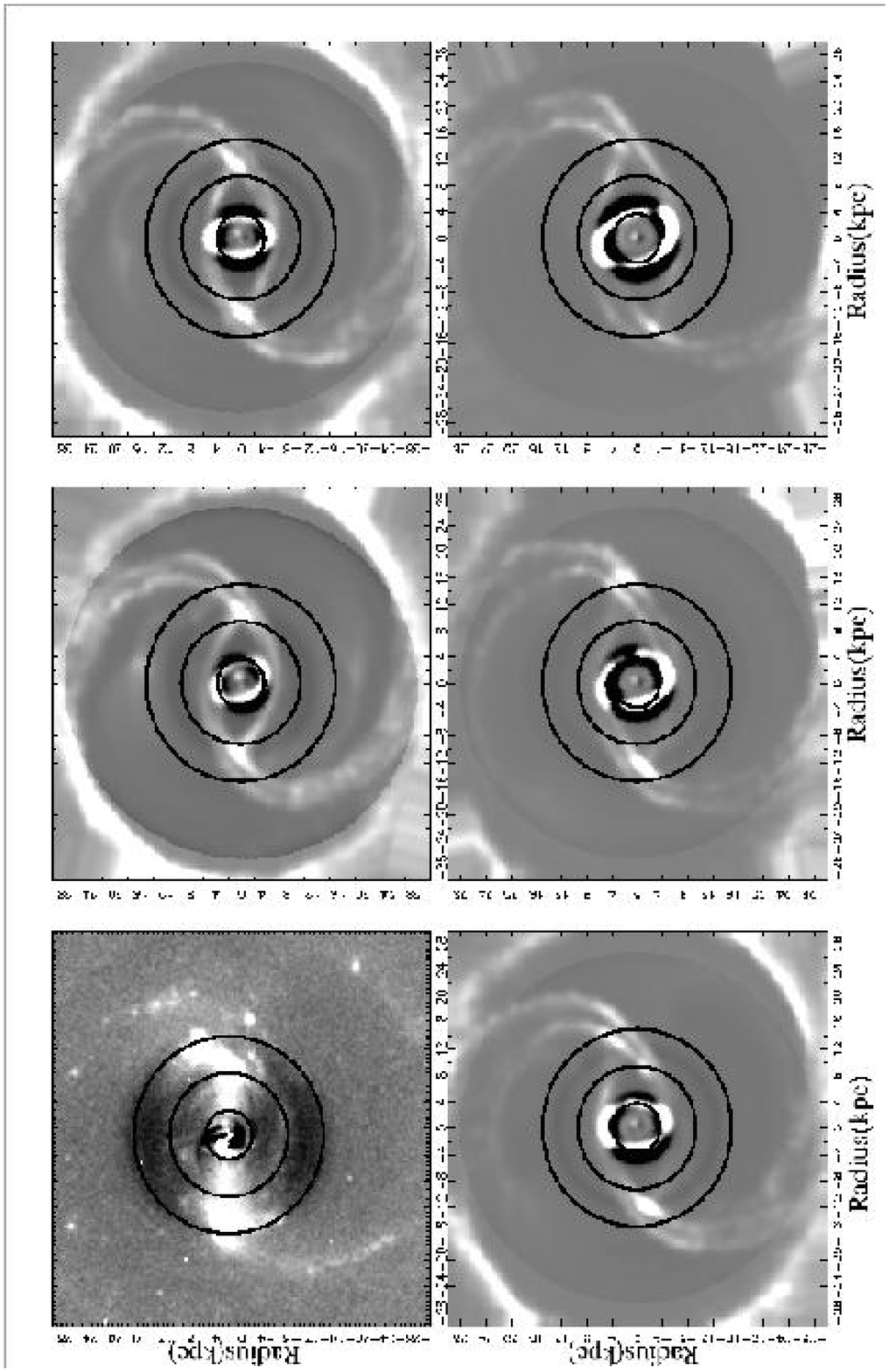,width=10.0cm,angle=-90} }}
\caption[Morphological comparison for NGC~7483]{The  top left figure
shows the masked $B$-band image of NGC~7483.  The top middle figure
shows the convolved and masked gas density map of the NGC~7483 model
with$R_{\rm CR}$/$R_{\rm bar}$=1.0, and the rest of the figures, from
left to right and top to bottom, the corresponding models with $R_{\rm
CR}$/$R_{\rm bar}$=1.2,1.3,1.4 and 1.6. The circles indicate the
position of the resonances for the axisymmetric approximation from
inside to outside, the ILR, IUHR and CR. In all figures, the
resonances are drawn for the $R_{\rm CR}$/$R_{\rm bar}$=1.0 case as a
reference.}
\label{fig:substract_NGC7483}
\end{center}
\end{figure*}

 A morphological comparison is being made for different pattern speeds
following the method explained in Section ~\ref{thebestfitmodels} (see
Fig.~\ref{fig:substract_NGC7483}). Both the nuclear and the inner
rings are present in the models. The size of these features when
compared to the real galaxy suggests a fast rotating bar.  At a larger
scale, the models show a two armed spiral pattern. In the simulations
some gas becomes trapped around the $L_{4,5}$ Lagrangian points with
the maximum density enhancement offset from the zero velocity points
. This effect can be seen in the density distribution
(Fig.~\ref{fig:ngc7483_vfield}).

\section{NGC 5505}
\label{ngc5505}

This is an isolated SBa galaxy at a distance of 57 Mpc.  There is no
kinematic nor photometric study of this galaxy in the
literature. Morphologically it is characterised by a  four arm spiral
structure and an inner ring. The bar is populated with HII regions and
the nucleus shows intense emission in H$_{\alpha}$. The bar position
angle is at about 40$^{\circ}$ with respect to the line of nodes.\\

\begin{figure*}
\begin{center}
\vspace{0cm}
\hspace{0cm}\psfig{figure=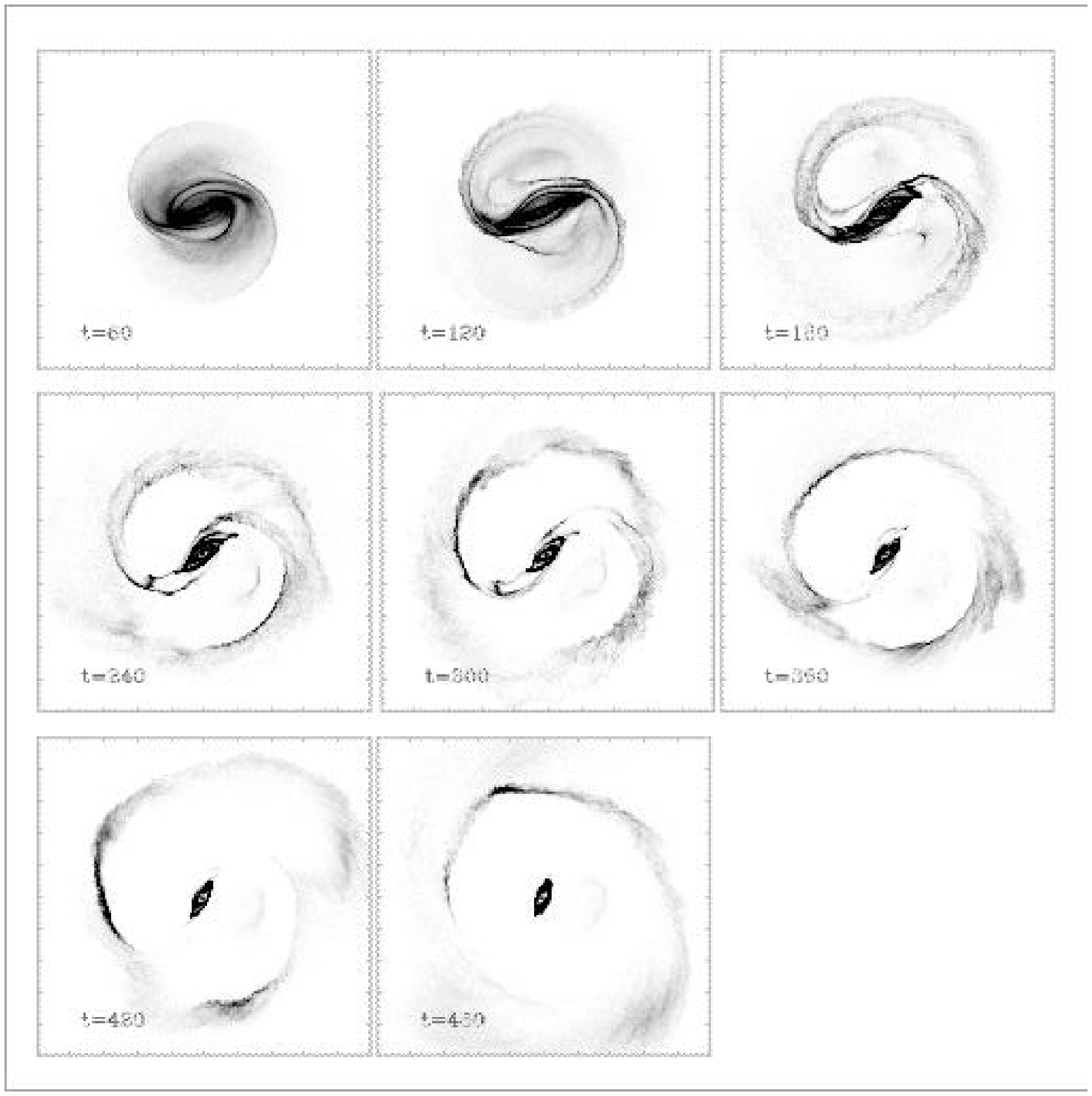,width=14.5cm,angle=-90}
\vspace{0cm}
\caption[Time evolution of the gas distribution of NGC5505]{Time
evolution of the gas distribution in the NGC5505 simulation with
corotation at 1.4$\times$the bar semi-major axis. Each panel, left to
right and top to bottom,  shows the distribution at different
integration times, from 1 bar rotation to 8 bar rotations, and in the
frame corotating with the bar. The non-axisymmetric component is fully
grown after three bar rotations and the bar pattern is rotating
counter-clockwise in the inertial frame. The size of the frames is 20
kpc in diameter and time is in Myr.}
\label{fig:ngc5505evolgas}
\end{center}
\end{figure*}

The gas was left to evolve for 8 bar rotations. The pattern speeds
explored in the simulations are shown in
Table~\ref{tab:gal_par}. NGC~5505 is one of the two galaxies for which
no steady state is reached for any of the pattern speeds explored. An
example of the time evolution of the gas distribution of NGC~5505 is
given in Fig.~\ref{fig:ngc5505evolgas} for a $R_{\rm CR}$/$R_{\rm
bar}$=1.0. The gas seems to dissipate energy through shocks and falls
into the centre. In the final configuration, most of the gas is
concentrated at the centre and some gas lies outside the
OLR. Fig.~\ref{fig:ngc5505evolgas} suggests that these shocks are due
to $x_{1}$ orbits that appear to loop over a large radial range. No
ILRs are present according to the frequency diagram in the
axisymmetric approximation, therefore the inner and nuclear ring do
not seem to be associated to such a resonance and instead they are
probably associated with the above $x_{1}$ orbits.\\

Since no steady configuration was found no comparison with the
observed data was carried out.  However, to test whether a stationary
state could be reached when including a rigid dark, runs with a dark
halo were carried out. The results are presented in
Section~\ref{NGC5505dm}.

\section{Models with a dark halo component}
\label{modelswithadarkhalocomponent}

We ran simulations with a dark halo component to test the effect of a
larger axisymmetric component on the gas distribution and the
kinematics of the modelled galaxies. Simulations were carried out on
two of the five galaxies; IC~5186 and NGC~5505.\\

The tests on NGC~5505 were done to check whether the introducing of a
dark halo could stabilise the gas flow. The dark halo was simulated by
letting the bar grow to a percentage of its onset time and then
continuing the simulations with this non-fully grown bar.   This
procedure mimics a lower stellar M/L, with the removed mass
redistributed in an axisymmetric dark component (see
Section~\ref{simulations}).  In this way we can test if the failure to
reach a steady state was due to the fact that they do have a dark
matter halo or because the modelling done here is not appropriate for
this galaxy (e.g. because the stellar distribution is not in a
sufficiently stationary state to allow a rigid potential
approximation). \\

This test was also done for IC~5186 (presented in the first part of
this Section), the galaxy on which we have done most of the test
simulations.

\subsection {IC 5186}
\label{IC5186dm}

The parameter space covered by the simulations is the same
as the one covered for the standard runs and three different dark halo
contribution to the total mass are tried, namely 5\% , 20\% and 40\%.\\

\begin{figure*}

\begin{center}
\centerline{\vbox{ \psfig{figure=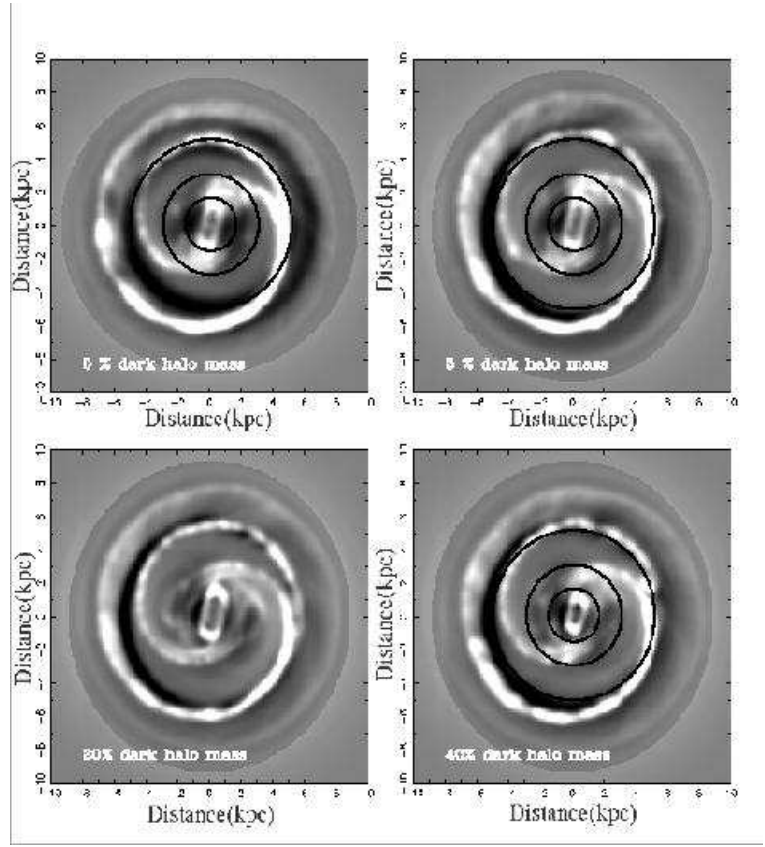,width=10.0cm} }}
\caption[Morphological comparison for IC~5186 with different dark halo
masses]{The top left figure shows the convolved and gas density map
for the standard IC 5186 model with the mass distribution derived from
the composite \HI-band and no dark halo.  The other figures show the
corresponding maps for the models with a 5\% dark halo mass (upper
right panel), 20\% dark halo mass (lower left panel), 40\% dark halo
mass (lower right panel). All models are taken at 8 bar rotations and
have $R_{\rm CR}$/$R_{\rm bar}$=1.4. The circles indicate the
resonances (IUHR, CR and OLR) in the axysimmetric approximation.}
\label{fig:mask_dm}
\end{center}
\end{figure*}

  There is a clear change in the gas distribution when increasing the
dark halo contribution, as one would expect from the increased
presence of the axisymmetric component in the potential. The gas
distribution is smoother and the inner ring seems to become more
circular. Outside the corotation radius the distribution is very
similar for all the dark halo mass percentages. One can see from
Fig.~\ref{fig:mask_dm} that there is no way to distinguish between the
0\% dark halo and the 5\% dark halo cases. However, for the cases with
20\% and 40\% dark halo mass, not only is the gas distribution very
different to the heavy disk cases but it also diverges from the light
distribution of the real galaxy significantly (see
Fig.~\ref{fig:light} and Fig.~\ref{fig:mask}).  The heavy disk models
show more similarities with the deprojected $B$-band image than the
light disk models.\\

\begin{figure*}

\begin{center}
\centerline{\vbox{ \psfig{figure=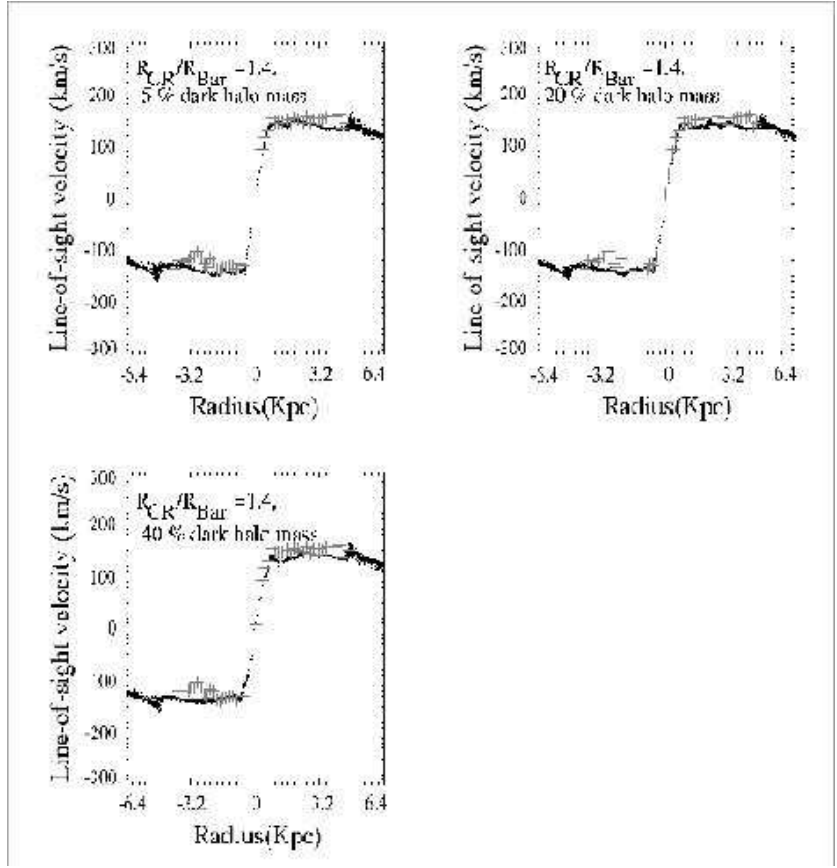,width=11.0cm} }}
\caption[Position-velocity diagrams along the major axis for the
IC~5186 models with $R_{\rm CR}$/$R_{\rm bar}$=1.4 and different dark
halo masses,  for different dark halo masses]{Line-of-sight velocity
curves along the major axis for the IC~5186 models with $R_{\rm
CR}$/$R_{\rm bar}$=1.4 and different dark halo masses, for different
dark halo masses, after 8 bar rotations. The dots represent the gas
particles and the overlaid gray crosses the observed major axis line-of-sight velocity curve. The dark halo masses percentages for each position-velocity
diagram are indicated in the panels. Notice that the fit to the
observed data worsens as the dark halo mass increases. For the
simulations with 40\% dark halo mass, the inner few arcsec show
velocity jumps not present in the observed kinematics. The
$\chi^{2}/N$ value becomes significantly larger than in the no halo
case only for the 40\% dark halo model.}
\label{fig:dm_rc}
\end{center}
\end{figure*}

The L-O-S velocity curves (see Fig.~\ref{fig:dm_rc}) show that
the agreement with the observed data worsens for the case with 40\%
dark halo mass. The $\chi^{2}/N$ value is 2.66 for the 5\% dark halo
mass, 2.69 for the 20\% dark halo mass and 3.79 for the case with 40\%
dark halo mass. The 5\% and the 20\% case are almost indistinguishable
from the case without a dark halo, whereas in the central 5 arcsec the
models with 40\% dark halo mass departs from the observed
 L-O-S velocity curve along the major axis.\\

The fact that a disk with 20\% of the mass residing in the dark halo
still gives a good agreement in the L-O-S velocity curve
comparison agrees with the definition of maximum disk given at the
introduction, where the stellar mass provides 85\%$\pm$10\% of the
total rotational support of the galaxy.\\

Runs with different pattern speeds ($R_{\rm CR}$/$R_{\rm bar}$=1.0 and
$R_{\rm CR}$/$R_{\rm bar}$=1.6) for the different dark halo masses
where also carried out. The $\chi^{2}/N$ values obtained for the
kinematic comparisons where always larger than the ones obtained for
$R_{\rm CR}$/$R_{\rm bar}$=1.4. The same behaviour as above is found
for $R_{\rm CR}$/$R_{\rm bar}$=1.0. However it is interesting to
notice that although the fit is not good for the case $R_{\rm
CR}$/$R_{\rm bar}$=1.6 the $\chi^{2}/N$ diminishes as the dark halo
mass increases.

\subsection {NGC 5505}
\label{NGC5505dm}

As seen in Section~\ref{ngc5505}, no stationary state was reached in
the simulations for NGC~5505 for any of the pattern speeds
explored. The experience with all the simulations suggests that a
slower bar would have produced an even more unstable result.  In order
to test if the gas could be stabilised with a dark halo component
simulations similar to those of the previous section on IC~5186 were
carried out.\\

For a minimum dark halo mass of 20\% the total mass a stationary state
is found for all pattern speeds. Fig.~\ref{fig:ngc5505_dm_rc} shows the
L-O-S velocity curve for NGC~5505 in these cases. No good
agreement with the data is found. Further discussion will be presented
in Section~\ref{doesngc5505needadarkhalo}.\\

 The fact is that the gas stabilises faster and evolves slower in the
simulations with a dark matter halo but it is possible that with
sufficient integration time the gas would have reached the same
configuration as the simulations without a dark halo. For a dark halo
mass of 5\%  the total mass the simulation evolves in a similar
fashion as in the case with no dark halo.

\section{Discussion}
\label{discussionchapter5}

\subsection{Why do some galaxies not reach a steady state?}

It is not clear why for some of the galaxies we obtained a stationary
configuration while for others (NGC~5505 and NGC~7267) we could not.
It is certainly related to the non-axisymmetric part of the potential,
since the final gas distribution in the latter cases is due to a high
degree of dissipation through the shocks. One could expect that in the
cases where the galaxies show more star forming regions the M/L
calculation is less accurate. However, no trend is found with colour
gradients. One of the galaxies with a larger colour change in the bar
and disk region is NGC~5505. However NGC~7483 also shows large colour
differences across the bar region (paper II) and NGC~7267 barely shows
any change, i.e. has an almost constant M/L value. See paper II for
the color maps of the individual galaxies. \\

In order to check the role played by light distribution asymmetries in
the galaxies, runs with an bisymmetrised mass distribution were
carried out for NGC~7267. Standard parameters were used for the
simulations. No steady configuration was reached after integrating for
8 bar rotations.\\

Since one expects a larger degree of dissipation in stronger bars the
bar strength in the sample galaxies was calculated to see if there was
any correlation. To calculate the bar strength one has to take into
account the radial and the tangential forces. The definition adopted
here for the bar strength \citep{combes81,block} is the maximum over
$R$ of

\begin{equation}
Q_T(R)=\frac{F^{\rm
max}_T(R)}{F_0(R)}=\frac{\frac{1}{R}|\frac{\partial\Phi(R,\phi)}{\partial
\phi}|_{\rm max}}{\frac{d\Phi_0(R)}{dR}}   \hspace*{1cm}
\end{equation}

where $F^{\rm max}_{T}(R)$ is the azimuthal maximum of the absolute
tangential force, $F_{0}(R)$ the mean axisymmetric radial force,
$\Phi(R,\phi)$ the in-plane gravitational potential and $\Phi_0(R)$
its axisymmetric part.\\

This way of measuring the bar strength takes into account the
axisymmetric disk in which the bar lies. Table~\ref{tab:strength}
shows the results of the bar strength for the modelled galaxies,
assuming the scale-heights adopted in the standard runs. In this
table, the bar class refers to the bar strength classification
introduced by \citet{block}, going from 1 to 6, with
class 6 representing the strongest bar. There seems to be no
correlation between bar strength and the steadiness of the
simulations. However, since the gas flow in the simulations for
NGC~5505 stabilises after adding a rigid halo, the steadiness of the
simulations must be related to the level of dissipation due to the
non-axisymmetric component.

\begin{table}
\begin{center}
\caption{Bar strength for the modelled galaxies.}
\label{tab:strength}
\vspace{5mm}
\begin{tabular}{clll}
\hline\hline Galaxy name& Bar strength& Bar class& Steady pattern\\ &
& & achieved\\  \hline NGC 7483&0.34&3& Yes\\ NGC 7267&0.56&6& No\\
NGC 5728&0.16&2& Yes\\ NGC 5505&0.44&4& No \\ IC 5186& 0.60&6& Yes\\
\hline\hline
\end{tabular}
\end{center}
\end{table}

\subsection{Does NGC~5505 need a dark halo?}
\label{doesngc5505needadarkhalo}

\begin{figure*}
\begin{center}
\centerline{\vbox{ \psfig{figure=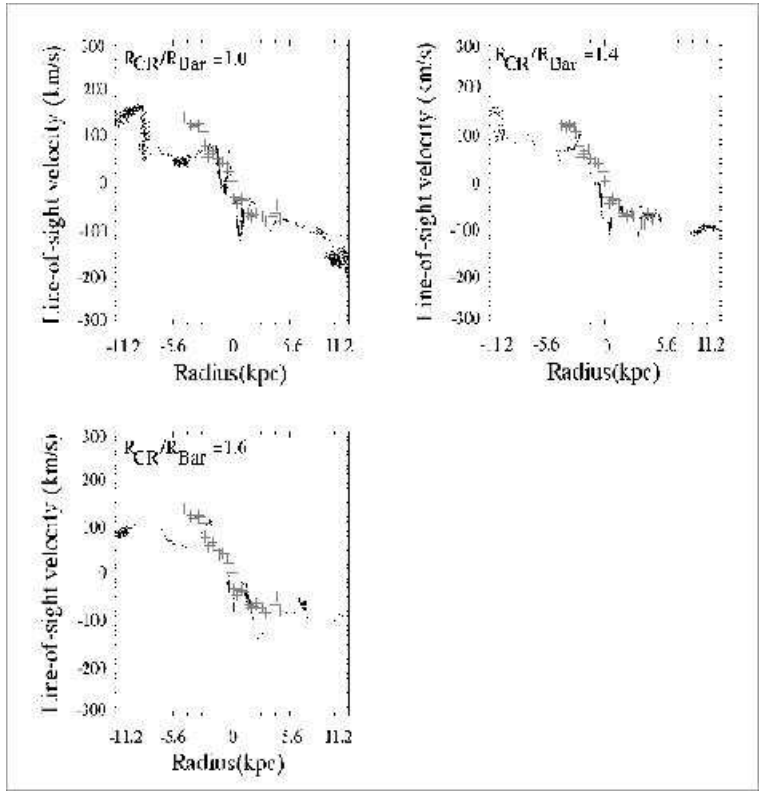,width=10cm} }}
\caption[Position-velocity diagrams along the major axis for models
with different bar pattern speeds for NGC~5505 models with a 20\% dark
halo component]{Line-of-sight velocity curves along the major axis for
the NGC 5505 models with different bar pattern speeds and a 20\% dark
halo component. The dots represent the gas particles and the overlaid
gray crosses the observed major axis line-of-sight velocity curve. From
left to right and top to bottom, $R_{\rm CR}$/$R_{\rm bar}$=1.0,1.4
and 1.6 respectively. No good fit is found for the models with dark
mass, and the models without a dark halo component do not reach a
steady state.}
\label{fig:ngc5505_dm_rc}
\end{center}
\end{figure*}

We cannot conclude with the present data anything about the dark
matter component in the inner parts of NGC~5505, until we understands
why the gas does not stabilise for NGC~5505 and NGC~7267. When one
adds a dark halo component the gas settles faster, however the
L-O-S velocity curve does not give a good fit to the data. It is
not possible to conclude anything about the need for a dark halo in
NGC~5505. All that has been shown is the relationship between the fact
that the gas does not reach a steady state and the addition of the
dark halo component whose effect is to reduce the radial gas flow and
smooth out the potential. The fact that the model velocities are too
low in the central regions (see Fig.~\ref{fig:ngc5505_dm_rc}) may be
suggesting that one should add a cuspy halo to increase the mass
density in that region. But also adding a cusp reduces the
non-circular motions and thus may lower the velocities in the L-O-S
velocity curve, depending on the orientation of the slit relative to the bar.
It is interesting to notice that, as pointed out by Athanassoula
(2001), a heavy halo can enhance the bar growth by the exchange in the
angular momentum between the two components, where the halo might take
angular momentum from the bar. She argues that the formation of a bar
causes the central concentration of the disk to increase so that the
disk dominates more and more in the inner region and  evolves into a
maximum disk after the bar has grown. Maybe NGC~5505 is an example of
a strong bar embedded in a heavy halo.

%

\section{Conclusions}
\label{conclusion}

 The aim of the modelling is to see whether the mass distribution
obtained from the light distribution in the $H$-band, together with
the hydrodynamical simulations, can reproduce the observed rotation
curves and the gas morphology of the inner regions of our
galaxies. The M/L is determined using population synthesis
models. Since no additional mass component is added, the M/L acts as a
normalisation factor on the rotation curves. Therefore, testing the
modelled L-O-S velocity curves is also a test of the population synthesis
models. This turns out to be largely a test of the adopted IMF (see
paper II). However, we are not only testing the normalisation factor
but also the shape of the central parts of the position-velocity
diagram which in the case of barred galaxies has the characteristic
imprints of the non-circular motions imposed by the bar. In this way
we can test whether or not a dark halo component is needed to explain
the observed kinematics in the inner regions of our galaxies.\\

The simulations presented here are compared to long slit
spectroscopy along the major axis, this could certainly produce
degeneracies in the obtained parameters. These galaxies were too faint
to obtain high spatial resolution HI maps and no optical integral
field unit was available at the time. In order to provide better
constraints, further comparison with 2-D velocity fields will be
necessary.\\

 Hydrodynamical simulations are run for five of the sample galaxies,
exploring different areas in parameter space. Different bar pattern
speeds and different scale-heights are investigated. For three of the
five galaxies a steady gas flow is reached after the non-axisymmetric
component is fully grown. For two of these galaxies (NGC~7483 and
IC~5186) the best fit pattern speed is found, both giving a fast bar
(with $R_{\rm CR}/R_{\rm bar}$=1.0 for NGC 7483 and $R_{\rm CR}/R_{\rm
bar}$=1.4 for IC~5186). For the third one, NGC~5728, the models do not
fit well the observed data, possibly due to the presence of a
secondary bar decoupled from the primary bar. However, the M/L ratio
adopted seems to give the right normalization factor. For the other
two galaxies (NGC~7267 and NGC~5505), no steady gas flow is achieved:
the dissipative shocks make the gas particles fall toward the centre,
leaving most of the gas at the centre and in an outer ring probably
associated to the OLR. However, when 20\% of the non-axisymmetric mass
component is replaced by an axisymmetric dark halo component then the
gas manages to reach a steady state.The simulations with short
scale-heights do not give a good fit for any of the pattern
speeds. However, simulations run with higher scale-heights give better
fits, although the scale-heights are unrealistic for real galaxies.\\

Simulations were run for potentials obtained from the
$I$-band light distribution and from a composite $H$-band image at the
centre and $I$-band in the outer isophotes. Such models show
differences in the gas distribution resulting from the higher dust
absorption in the central regions of the $I$-band image.  For example,
the simulations in the potential derived from the $I$-band image alone
show a substantially more pronounced nuclear ring, whereas the
simulations in the potential derived from the composite $(H\!+\!I)$
image seem to have more gas trapped around the $L_{4/5}$ Lagrangian
points. However, only small differences between the two simulations
are found in the kinematics along the major axis.\\

 We tested the effect of a dark halo component in the simulations,
by converting part of the visible mass inferred from the population synthesis models into an axisymmetric component, simulating the effect of a lower stellar M/L and rigid halo. The fit to the observations worsens significantly.\\

The M/L ratios is obtained from the population synthesis models give the right
normalisation factor in order to match the modelled position-velocity diagrams
to the observed rotation curve, indicating that the population synthesis
models do give realistic M/L$_{\rm H}$ values. For the above galaxies in which
full comparison of the rotation curve has been done, not only the
normalisation, but the shape of the rotation curve agrees very well with the
data, even in the case of NGC 7483 which possesses a very complicated and messy
rotation curve. 

\begin{acknowledgements}
 We would like to thank the APAC super computer facility for providing
 the computing time and facilities that made this work possible. This
 work was supported by the Australian National University through an
 ANU PhD. grant. We would also like to thank the external referees of
 the PhD. thesis associated to this paper for their useful comments
 and Herv\'e Wozniak for his constructive criticism; which helped to the
 improvement of this manuscript.
\end{acknowledgements}

\begin{thebibliography}{28}
\expandafter\ifx\csname natexlab\endcsname\relax\def\natexlab#1{#1}\fi

\bibitem[{Athanassoula(1992)}]{athanassoula92}
Athanassoula, E. 1992, MNRAS, 259, 345

\bibitem[{Athanassoula(2002)}]{athanassoula02}
Athanassoula, E. 2002, ApJ, 569, 83

\bibitem[{Benz(1990)}]{benz90}
Benz, W. 1990, in NATO ASI Ser. C, Vol. 302, The Numerical Modelling of
  Nonlinear Stellar Pulsations Problems and Prospects (Kluwer), 269

\bibitem[{Block {et~al.}(2002)Block, Buta, Puerari, Knapen, Elmegreen, Stedman,
  \& Elmegreen}]{block}
Block, D., Buta, R., Puerari, I., {et~al.} 2002, in ASP Conference Proceedings,
  Vol. 273, The Dynamics, Structure and History of Galaxies, 97

\bibitem[{Carter(1995)}]{carter}
Carter, B. 1995, Ap\&SS, 230, 163

\bibitem[{Combes(1994)}]{combes94}
Combes, F. 1994, in Mass-Transfer Induced Activity in Galaxies, ed. Shlosman,
  Vol. 170

\bibitem[{Combes \& Leon(2002)}]{combes02}
Combes, F. \& Leon, S. 2002, in EdP-Sciences (Editions de Physique), Vol. 379,
  SF2A-2002: Semaine de l'Astrophysique Francais, ed. F.~Combes \& D.~Barret,
  403

\bibitem[{Combes \& Sanders(1981)}]{combes81}
Combes, F. \& Sanders, R. 1981, A\&A, 96, 164

\bibitem[{Courteau \& Rix(1996)}]{courteau99}
Courteau, S. \& Rix, H.-W. 1996, ApJ, 513, 561

\bibitem[{Crocker {et~al.}(1996)Crocker, Baugus, \& Buta}]{crocker}
Crocker, D., Baugus, P., \& Buta, R. 1996, in PASP, Vol.~91, Barred galaxies,
  ed. R.~Buta, D.~Crocker, \& B.~Elmegreen, 80

\bibitem[{de~Jong(1996)}]{jong96}
de~Jong, R.~S. 1996, A\&A, 313, 45

\bibitem[{Debattista \& Sellwood(1998)}]{debattista98}
Debattista, V. \& Sellwood, J. 1998, ApJ, 493, 5

\bibitem[{Freeman(1992)}]{freeman92}
Freeman, K. 1992, IAUS, 149, 65

\bibitem[{Freudenreich(1998)}]{freudenreich98}
Freudenreich, H. 1998, ApJ, 492, 495

\bibitem[{Fux(1997)}]{fux97}
Fux, R. 1997, PhD thesis, Geneva University

\bibitem[{Fux(1999)}]{fux99}
Fux, R. 1999, A\&A, 345, 787

\bibitem[{Kregel {et~al.}(2002)Kregel, der Kruit, \& de~Grijs}]{kregel02}
Kregel, M., der Kruit, P., \& de~Grijs, R. 2002, MNRAS, 334, 646

\bibitem[{Landolt(1992)}]{landolt}
Landolt, A. 1992, AJ, 107, 372

\bibitem[{P\'erez(2003)}]{perez3}
P\'erez, I. 2003, PhD thesis, Australian National University

\bibitem[{Persson {et~al.}(1998)Persson, Murphy, Krzeminski, Roth, \&
  Rieke}]{persson}
Persson, S., Murphy, D., Krzeminski, W., Roth, M., \& Rieke, M. 1998, AJ, 116,
  2475

\bibitem[{Pfenniger \& Friedli(1993)}]{pfenniger93}
Pfenniger, D. \& Friedli, D. 1993, A\&A, 270, 561

\bibitem[{Regan \& Teuben(2003)}]{regan}
Regan, M. \& Teuben, P. 2003, ApJ, 582, 723

\bibitem[{Sackett(1997)}]{sackett97}
Sackett, P. 1997, ApJ, 483, 103

\bibitem[{Salucci \& Persic(1999)}]{salucci99}
Salucci, P. \& Persic, M. 1999, A\&A, 351, 442

\bibitem[{Tremaine \& Ostriker(1999)}]{tremaine}
Tremaine, S. \& Ostriker, J. 1999, MNRAS, 306, 662

\bibitem[{Weinberg(1985)}]{weinberg85}
Weinberg, M. 1985, MNRAS, 213, 451

\bibitem[{Weiner {et~al.}(2001)Weiner, Sellwood, \& Williams}]{weiner01}
Weiner, B., Sellwood, J., \& Williams, T. 2001, ApJ, 546, 931

\bibitem[{Wozniak {et~al.}(1995)Wozniak, Friedli, Martinet, Martin, \&
  Bratschi}]{wozniak95}
Wozniak, H., Friedli, D., Martinet, L., Martin, P., \& Bratschi, P. 1995, A\&A,
  111, 115

\end{thebibliography}

\newpage
\begin{table}
\begin{center}
\caption{Model parameters for the modelled galaxies}
\label{tab:gal_par}
\vspace{5mm}
\begin{tabular}{llccc}
\hline\hline 
Galaxy name& $R_{\rm CR}$/$R_{\rm bar}$& Corotation& $\Omega_{\rm b}$
& $\chi^{2}/N$\\ & &(kpc)&(km s$^{-1}$ kpc$^{-1}$)& \\ 
\hline 
IC 5186&1.00& 2.2&86.2&6.69\\ 
&1.10 & 2.4 & 79.6&5.94\\ 
&1.20 & 2.6 & 73.6&2.36\\ 
&1.30 &2.8 & 68.3&2.55\\ 
&1.40 & 3.0 & 63.4&2.25\\ 
&1.60 & 3.5 & 55.2&3.56\\
\hline
NGC 5728&1.00& 5.1 & 81.9\\
&1.10&5.6& 73.3&\\ 
&1.20& 6.2& 66.5&\\ 
&1.30&6.7&60.9&\\ 
&1.40&7.2& 56.2&\\
&1.60&8.2&48.9&\\ 
\hline
NGC 7267&1.00 &4.5& 44.8&\\ 
&1.10& 4.9& 40.7&\\ 
&1.20& 5.4& 37.2&\\ 
&1.30& 5.8& 34.1&\\ 
&1.40& 6.3& 31.4&\\
&1.60& 7.2& 27.1&\\ 
\hline
NGC 7483 &1.00&9.6 &27.5&1.02\\ 
&1.10& 10.5 &24.8&2.25\\ 
&1.20 &11.5& 22.7&2.04\\ 
&1.30&12.4& 20.8&2.05\\ 
&1.40& 13.4&19.2&1.65\\ 
&1.60& 15.3& 16.3&1.87\\
\hline
NGC 5505&1.00& 3.3& 99.3&\\ 
&1.10&3.7&91.2\\ 
&1.20 &4.0 &83.2&\\ 
&1.30& 4.3& 75.8&\\ 
&1.40& 4.6& 69.3&\\ 
&1.60&5.3 &59.1&\\ 
\hline\hline
\end{tabular}
\end{center}
\end{table}

\begin{figure*}
\begin{center}
\centerline{\vbox{ \psfig{figure=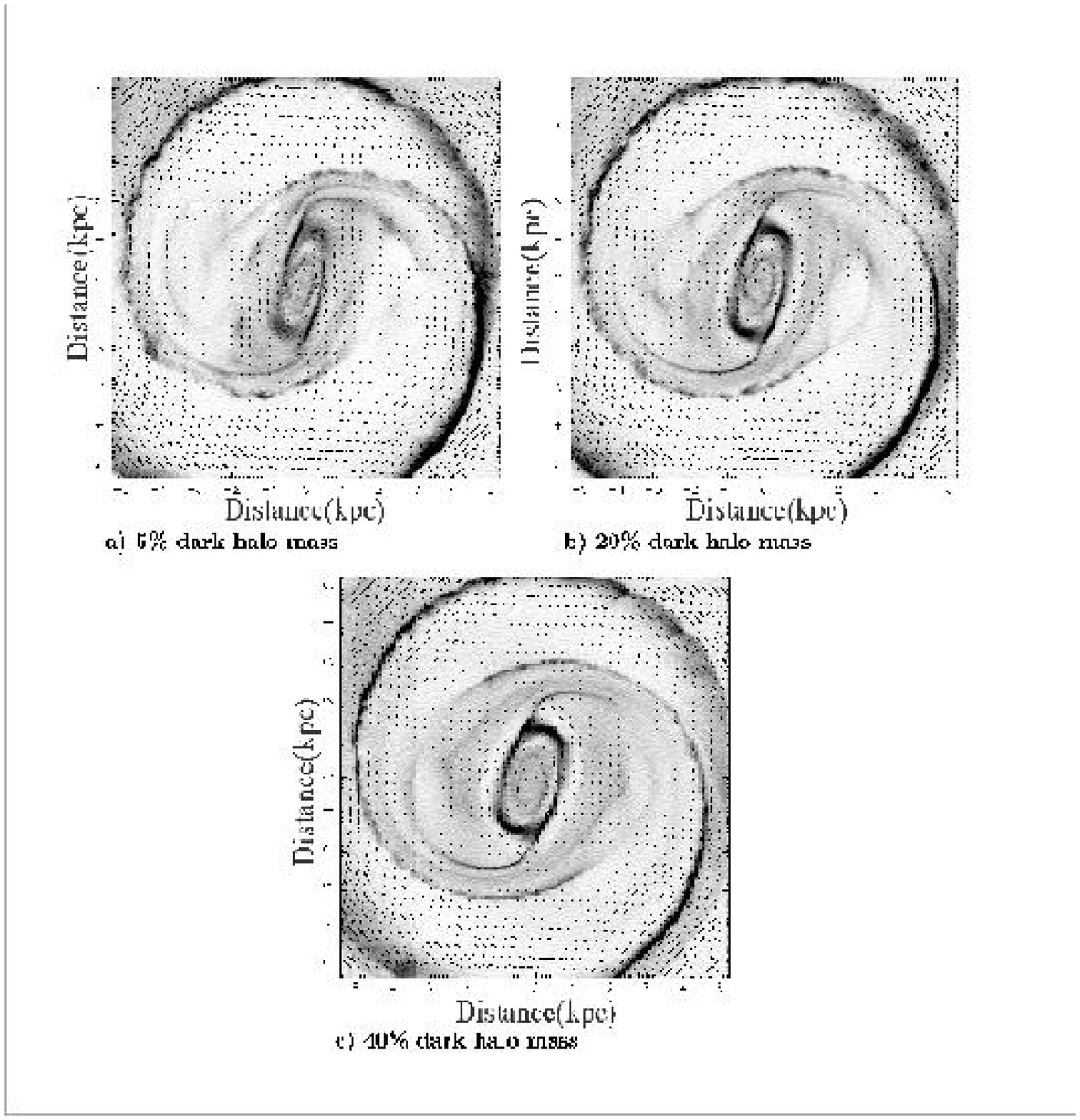,width=15cm} }}
\caption[Velocity field of IC~5186 with different dark halo
masses]{The gas distribution and velocity field for the mass
distributions of IC 5186 derived from the composite \HI-band image and
with various dark halo mass fractions, at 8 bar rotations and for
$R_{\rm CR}$/$R_{\rm bar}$=1.4. The length of the vectors is
proportional to the velocity in the frame rotating with the bar.}
\label{fig:dm}
\end{center}
\end{figure*}

\begin{figure*}[h]
\begin{center}
\centerline{\vbox{ \psfig{figure=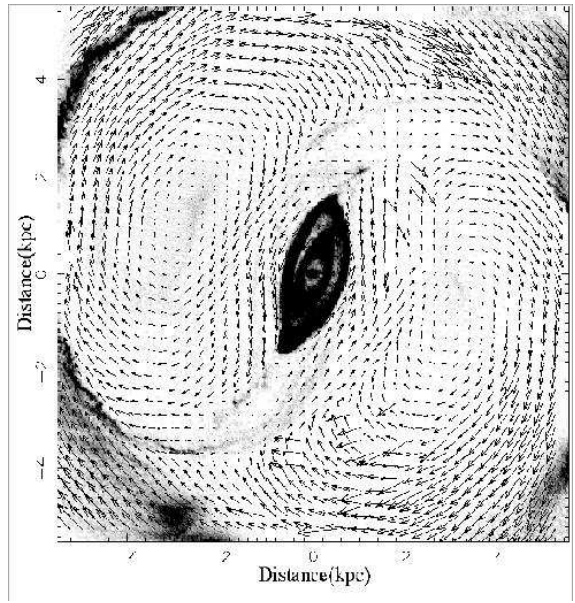,width=7.5cm}
}}
\caption[The gas distribution and velocity field for the mass derived
from the composite \HI-band image plus a 20\% dark halo mass of
NGC~5505]{The gas distribution and velocity field for the mass
distribution of NGC 5505 derived from the composite \HI-band image and
with 20\% dark halo, at 8 bar rotations and for $R_{\rm CR}$/$R_{\rm
bar}$=1.0.  The length of the vectors is proportional to the velocity
in the frame rotating with the bar.}
\label{fig:vfield_NGC5505}
\end{center}
\end{figure*}

\end{document}